\documentclass[manuscript]{acmart}
\usepackage{threeparttable}
\usepackage{booktabs}
\usepackage{tabularx}
\usepackage{multirow}
\usepackage{tablefootnote}
\usepackage{threeparttable}
\usepackage{pifont}
\usepackage[skins]{tcolorbox}
\tcbuselibrary{breakable}
\usepackage[linesnumbered,ruled,vlined,noend]{algorithm2e}
\usepackage{caption}
\usepackage{listings}
\usepackage{subcaption}
\usepackage{xurl}
\usepackage{enumitem}
\usepackage{graphicx}
\usepackage{bbding}
\usepackage{amsmath}
\usepackage{amsthm}
\usepackage{siunitx}
\usepackage{makecell}
\usepackage{wrapfig}
\newtheoremstyle{mydef} 
  {5pt}{5pt}             
  {}                     
  {}                     
  {\bfseries}            
  {.}                    
  {0.5em}                
  {}                     

\theoremstyle{mydef}
\newtheorem{definition}{Definition}[section] 

\newtcolorbox{promptbox}[1][]{
    colback=gray!10,          
    colframe=gray!50,         
    fonttitle=\bfseries,      
    coltitle=gray!40!black,   
    title=Prompt,             
    breakable,                
    enhanced,
    sharp corners,
    boxrule=0.5pt,            
    boxsep=2mm,               
    left=0pt, right=0pt,      
    before=\vspace{2mm}, after=\vspace{2mm},
    #1
}

\makeatletter
\patchcmd\algocf@Vline{\vrule}{\vrule \kern-0.4pt}{}{}
\patchcmd\algocf@Vsline{\vrule}{\vrule \kern-0.4pt}{}{}
\makeatother

\SetCommentSty{mycommfont}
\SetKwComment{Comment}{\ \#\ }{}
\SetKwProg{Func}{function}{}{end}

\tcbset{
  my box2/.style={
    enhanced,
    colframe=#1!80,
    colback=#1!5,
  },
}
\newtcolorbox{summary-rq}{
  my box2=black,
  boxrule=1pt,top=3pt,bottom=3pt,left=4pt,right=4pt
}

\AtBeginDocument{%
  }

\setcopyright{acmlicensed}
\copyrightyear{2026}
\acmYear{2026}
\acmDOI{XXXXXXX.XXXXXXX}
\acmConference[Conference acronym 'XX]{Make sure to enter the correct
  conference title from your rights confirmation email}{June 03--05,
  2026}{Woodstock, NY}
\acmISBN{978-1-4503-XXXX-X/2026/06}

\begin{document}

\title{OODEval: Evaluating Large Language Models on Object-Oriented Design}


\author{Bingxu Xiao}
\orcid{0009-0005-3960-9248}
\affiliation{%
  \institution{Northwestern Polytechnical University}
  \city{Xi'an}
  \country{China}
}
\email{bingxuxiao@mail.nwpu.edu.cn}

\author{Yunwei Dong}
\orcid{0000-0001-9882-9121}
\authornotemark[0]
\authornote{Corresponding author.}
\affiliation{%
  \institution{Northwestern Polytechnical University}
  \city{Xi'an}
  \country{China}
}
\email{yunweidong@nwpu.edu.cn}

\author{Yiqi Tang}
\orcid{0009-0001-7884-9042}
\affiliation{%
  \institution{Northwestern Polytechnical University}
  \city{Xi'an}
  \country{China}
}
\email{atcraft@mail.nwpu.edu.cn} 

\author{Manqing Zhang}
\orcid{0000-0001-9086-0503} 
\affiliation{%
  \institution{Northwestern Polytechnical University}
  \city{Xi'an}
  \country{China}
}
\email{zmqgeek@mail.nwpu.edu.cn} 

\author{Yifan Zhou}
\orcid{0009-0005-1641-6590} 
\affiliation{%
  \institution{Southern University of Science and Technology}
  \city{Shenzhen}
  \country{China}
}
\email{12332419@mail.sustech.edu.cn} 

\author{Chunyan Ma}
\orcid{0000-0002-7549-7926} 
\affiliation{%
  \institution{Northwestern Polytechnical University}
  \city{Xi'an}
  \country{China}
}
\email{machunyan@nwpu.edu.cn} 

\author{Yepang Liu}
\orcid{0000-0001-8147-8126} 
\affiliation{%
  \institution{Southern University of Science and Technology}
  \city{Shenzhen}
  \country{China}
}
\email{liuyp1@sustech.edu.cn}

\renewcommand{\shortauthors}{Xiao et al.}

\begin{abstract}
Recent advances in large language models (LLMs) have prompted extensive evaluations throughout software engineering tasks, yet existing studies primarily focus on code-level performance, leaving LLMs’ capabilities in software design underexplored. This gap stems mainly from the absence of standardized benchmarks and automatic evaluation metrics in this domain. To fill these gaps, this work introduces a novel evaluation framework for object-oriented design (OOD) tasks, comprising new benchmarks and metrics. Specifically, we first present OODEval, a manually constructed benchmark covering 50 OOD  tasks categorized into three difficulty levels. To facilitate comparisons with human performance and the evaluation of our new metrics, we introduce OODEval-Human, the first human-rated OOD benchmark with 940 undergraduate-submitted solutions with instructor ratings. We further propose CLUE (Class Likeness Unified Evaluation), an automatic metric set for assessing global and fine-grained OOD performance. Leveraging this framework, we conduct a comprehensive empirical study evaluating 29 LLMs on OOD tasks, investigating five research questions: overall correctness, comparison with humans, model dimension analysis, task feature analysis, and bad case analysis. Results reveal that LLMs attain high syntactic accuracy but suffer from semantic deficiencies, particularly in method and relationship generation. Top-performing LLMs approach average undergraduate levels but lag behind expert humans. Key influencers include parameter scale, code specialization, and instruction tuning, while design complexity and requirement readability hinder performance. Common failure modes involve omissions and inaccuracies in classes, methods, and relationships. Overall, this study establishes a systematic evaluation framework for OOD tasks and offers empirical insights to inform future research, deployment, and education on LLMs in software design.
\end{abstract}




\maketitle

\section{Introduction}

In recent years, large language models (LLMs) such as GPT\cite{achiam2023gpt} and LLaMA\cite{touvron2023llama} have made remarkable advancements in the field of software engineering. To systematically assess their practical effectiveness, researchers have developed various benchmarks~\cite{10.1145/3597503.3639219,10.1145/3691620.3695470,11029911,10.1145/3691620.3695513} for software engineering tasks and conducted empirical studies to analyze the strengths and limitations of LLMs. However, recent studies~\cite{10.1145/3786771} indicate that existing benchmarks and empirical research are mainly focus on programming assistance and quality management tasks (77\%), while requirements and design tasks represent only 8\%. Moreover, the limited requirements and design tasks researches only focus primarily on isolated scenarios of requirements analysis and verification, which lacks systematic evaluation of LLMs’ capabilities from requirements to design, thus leaving it unclear how LLMs perform in software design scenarios.

To address this gap, this study focuses on object-oriented design (OOD)~\cite{10.1145/989791.989795} tasks, investigating LLMs’ ability to analyze natural language requirements and generate class diagrams. OOD plays a crucial role in ensuring software quality and reducing communication overhead within development teams. Therefore, systematically evaluating LLMs’ performance on OOD tasks holds both academic significance and practical relevance, shedding light on their potential applications in the software design phase. 

\textbf{Existing OOD Research Limitations.} To conduct such a systematic evaluation, it is necessary to establish standardized benchmark datasets and evaluation metrics. However, our review of research over the past decade indicates a lack of widely accepted benchmark datasets for OOD and unified automatic evaluation metrics. \textbf{\textit{[OOD Benchmark Limitation.]}} Table~\ref{tab:datasets_over_past_decade} summarizes the datasets used in related studies. Our findings reveal several limitations in existing OOD datasets: (1) most studies rely on case studies of individual examples rather than empirical evaluations on standard benchmarks; (2) most such cases or datasets are not publicly available; (3) some publicly available datasets are presented as images, which hinders automated evaluation; and (4) existing datasets or cases are relatively simple and lack stratification based on difficulty. \textbf{\textit{[Evaluation Metrics Limitation.]}} Next, we conducted a survey of existing evaluation methods and metrics for assessing the correctness of class diagram designs based on reference designs, with representative classical methods summarized in Table~\ref{tab:metric_survey}. Our findings indicate that traditional text-matching metrics, such as BLEU\cite{papineni2002bleu}, fail to capture the structural and semantic information in class diagrams. Embedding-based metrics, like BertScore\cite{Zhang*2020BERTScore:} and CodeBertScore\cite{zhou-etal-2023-codebertscore}, effectively capture overall semantics but often overlook structural details and fine-grained local semantics. Although domain-specific similarity evaluation methods have been proposed\cite{6748477,6933505,nikiforova2015approach,yuan2020structural}, they exhibit several limitations: (1) lack of reproducible open-source code for evaluation algorithms; (2) failure to balance global performance evaluation with local performance assessment of class attributes, methods, and relationships; and (3) insufficient utilization of structural and semantic information, neglecting fine-grained details.

\textbf{Novel Benchmarks and Evaluation Metric.} To address the limitations of existing benchmarks in OOD evaluation, we manually constructed a model-level benchmark dataset, OODEval, which comprises 50 tasks from requirements to design and encompasses three difficulty levels: simple, moderate, and hard. As shown in Table~\ref{tab:datasets_over_past_decade}, OODEval surpasses previous datasets in scale and diversity, offering PlantUML code formats for automated evaluation and visualized formats for human review. Furthermore, we created OODEval-Human, the first human-rated OOD benchmark, containing 940 undergraduate-submitted solutions with instructor ratings. This benchmark was constructed for two primary purposes: (1) to investigate the performance gap between LLMs and humans, and (2) to measure the correlation between newly proposed evaluation metrics and human ratings, which is a widely used approach for assessing the performance of new evaluation metrics~\cite{zhou-etal-2023-codebertscore,zhuo-2024-ice}. To overcome the limitations of existing evaluation metrics, we propose the CLUE (Class Likeness Unified Evaluation) metrics, which includes an overall performance metric along with four local performance metrics. This metric integrates more complete structural and semantic information from class diagrams to evaluate design correctness. We evaluate the performance of CLUE metrics by measuring the correlation between CLUE scores and human ratings on OODEval-Human. The results indicate a strong positive correlation, with a Pearson coefficient of 0.59, and demonstrate that CLUE’s evaluations align closely with human preferences.

\textbf{Empirical study and Findings.} To assess the capability of LLMs in OOD, we conducted the first comprehensive empirical study on 29 diverse LLMs using our constructed benchmark and evaluation metrics. \textbf{\textit{[Research Questions.]}} We set five research questions to examine the overall performance of LLMs, their differences compared to human undergraduates, the influence of model dimensions and task features on model performance, and the failure modes arising in the generation process of LLMs. \textbf{\textit{[Key Findings.]}} Our key findings are as follows: (1) LLMs exhibit strong syntactic correctness but show significant weaknesses in semantic correctness. (2) LLMs present weaker performance in generating class methods and relationships, with pronounced deficiencies in the hard subset, highlighting a clear direction for future LLM enhancements. (3) Qwen3-Coder-30B achieves top-tier results, rivaling online models like DeepSeek-R1 and GPT-4o. Notably, the Gemma3-4B-IT, the smallest in size, surpasses most local models and even GPT-4o-mini, informing model selection under resource constraints. (4) The state-of-the-art LLMs approach the average performance of humans, underscoring the need for software engineering education to implement safeguards against AI-assisted cheating and academic misconduct. Nevertheless, LLMs perform substantially below the best humans. (5) Parameter scale, code specialization, and instruction tuning emerge as key factors influencing OOD task performance. (6) Task features, including the number of classes, methods, relationships, and requirement readability, significantly influence LLM performance, indicating a need for future benchmarks to incorporate more complex designs in these dimensions. (7) Common failure modes of LLMs include keyword errors, missing classes and associations, and method omissions, which are observed across model families. However, the DeepSeek series, GPT series, Gemma3-IT, and Qwen3-Coder-30B demonstrate robust performance, infrequently exhibiting these issues. In summary, these findings inform LLM optimization, model selection, benchmark development, and educational strategies in software engineering.

In summary, this paper makes the following contributions:

\begin{itemize}
    \item \textbf{OODEval Benchmark}: A novel model-level benchmark for evaluating OOD generation, which is manually constructed  and publicly available, including 50 tasks from requirements to class diagram designs and spans three difficulty levels—simple, moderate, and hard.
    \item \textbf{OODEval-Human Dataset}: The first human-rated OOD dataset, which includes 940 undergraduate-submitted solutions along with instructor ratings for each submission. This dataset enables comparisons of OOD capabilities between LLMs and humans and supports the evaluation of correlations between new evaluation metrics and human assessments.
    \item \textbf{CLUE Metrics}: A novel OOD automatic evaluation metric set for assessing global and fine-grained OOD performance. These metrics are confirmed to be high correlation with human ratings on OODEval-Human.
    \item \textbf{Empirical Study}: A comprehensive evaluation of 29 diverse LLMs and human undergraduates on OOD tasks using OODEval and OODEval-Human. Our findings provide insights into model capabilities and highlight future research directions for LLMs on OOD tasks.
\end{itemize}

\section{Related Works}

\subsection{Existing Datasets for OOD Generation}
 
OOD~\cite{10.1145/989791.989795} refers to a task analyzing and designing class diagrams from requirements. We reviewed studies since 2010 on the generation of class diagrams from natural language requirements and summarized the datasets within these studies in Table~\ref{tab:datasets_over_past_decade}. Most studies use case studies instead of standardized datasets, limiting reproducibility and generalizability. Moreover, most datasets and case studies are not publicly available. After excluding non-public datasets and case studies, only two works remain relevant:~\cite{shweta-etal-2023-advancing} and~\cite{10.1145/3674805.3690741}. The dataset in~\cite{shweta-etal-2023-advancing}, a fragment-level dataset constructed via crowdsourcing, contains 63 design fragments paired with requirement snippets. However, it lacks complete requirements and class diagrams, making it unsuitable for evaluating end-to-end OOD generation. In contrast,~\cite{10.1145/3674805.3690741} provides a model-level dataset with 20 complete requirements and class diagrams. However, the dataset's ability to evaluate LLMs in complex design scenarios is limited by its overall simplicity and lack of diversity. According to the OODEval rating method, it includes only 15 simple and 5 moderate instances, which also lack variety in key areas such as class methods and requirement context length. To address the limited complexity and diversity of existing datasets, we developed OODEval, a comprehensive dataset with 50 samples across three difficulty levels (18 simple, 20 moderate, 12 hard; see Table~\ref{tab:oodeval-characteristics}). Compared to~\cite{10.1145/3674805.3690741}, OODEval has higher mean and standard deviation for class count, attribute count, method count, relationship count, and requirement context length. These metrics show that OODEval captures greater design complexity and requirements variety, allowing a more robust evaluation of the capabilities of LLMs in the generation of OOD.

\begin{table}[h]
\vspace{-3mm}
\setlength{\tabcolsep}{4.2pt}
\caption{Existing Datasets for OOD Generation.}
\vspace{-2mm}
\label{tab:datasets_over_past_decade}
\resizebox{\textwidth}{!}{
{\small
\begin{tabular}{llclcccccccll}
\toprule
& \textbf{Source} & \textbf{Year} & \textbf{Granularity} & \textbf{\# Data} & \textbf{\# Class} & \textbf{\# Attribute} & \textbf{\# Method} & \textbf{\# Relation} & \textbf{\# Word} & \textbf{Public?} & \textbf{Format} & \\ 
\midrule 
& Krishnan and Samuel\cite{5670730} & 2010 & Model-level  & 1  & - & - & - & - & - & \XSolidBrush  & Case   &\\
& Bajwa and Choudhary\cite{bajwa2011natural} & 2012 & Model-level & 5  & -  & - & -  & - & - & \XSolidBrush & Case &\\
& Landhäußer et al.\cite{landhausser2014requirements}& 2014 & Model-level & 7 & -  & -  & - & - & 140 $\pm$ 62  & \CheckmarkBold &  Case  &\\
& Sharma et al.\cite{7337625}& 2015 & Model-level & 5 & 11 $\pm$ 3  & 2 $\pm$ 2 & -  & -  & - & \XSolidBrush & Case &\\
& Ben Abdessalem Karaa et al. \cite{ben2016automatic} & 2016 & Model-level  & 1  & -  & -  & -  & - & - & \XSolidBrush  & Case  &\\
& Abdelnabi et al. \cite{9329301} & 2020 & Model-level  & - & -  & - & - & - & -  & \XSolidBrush & -   &\\
& Alharbia et al. \cite{alharbia2021framework}& 2021 & Model-level  & 3 & - & -  & - & - & - & \CheckmarkBold & Case &\\
& Bashir et al. \cite{9428817} & 2021 & Model-level & 3  & - & - & -  & - & - & \XSolidBrush & Case &\\
& Yang and Sahraoui\cite{10.1145/3550356.3561592}& 2022 & Fragment-level & 63 & 5 $\pm$ 2 & - & - & 5 $\pm$ 3 & 105 $\pm$ 54 & \CheckmarkBold & PNG/TXT &\\
& X et al. \cite{shweta-etal-2023-advancing}& 2023 & Model-level & 32 & -  & - & - & - & - & \XSolidBrush & XML  &\\
& Meng and Ban \cite{meng2024automated}& 2024 & Fragment-level  & 100 & 4 & - & - & 4 & - & \XSolidBrush & -  &\\
& Babaalla et al. \cite{10.1145/3659677.3659742}& 2024 & Model-level & -  & - & - & - & - & - & \XSolidBrush & -  &\\
& De Bari et al. \cite{10.1145/3674805.3690741}& 2024 & Model-level & 20 & 7 $\pm$ 2 & 13 $\pm$ 8 & 2 $\pm$ 5  & 8 $\pm$ 2 & 184 $\pm$ 58 &  \CheckmarkBold & PNG &\\
\midrule
& \textit{OODEval} & 2026 & Model-level & 50  & 8 $\pm$ 3  & 19 $\pm$ 10  & 17 $\pm$ 16 & 9 $\pm$ 4 & 237 $\pm$ 101 &  \CheckmarkBold & JSON/PNG &\\ 
\bottomrule
\end{tabular}
}
}
\vspace{-2mm}
\end{table}

\begin{table}[h]
\setlength{\tabcolsep}{4.2pt}
\caption{Existing Evaluation Methods for OOD Generation.}
\vspace{-2mm}
\label{tab:metric_survey}
\resizebox{\textwidth}{!}{
{\small
\begin{tabular}{clcllccccccc}
\toprule
& \textbf{Source} & \textbf{Year} & \textbf{Metric Category} & \textbf{Method Category} & \textbf{Automatic?} & \textbf{CStS} & \textbf{CSmS} & \textbf{SCM} & \textbf{CEMCS} & \textbf{Public?} &\\ 
\midrule 
& BLEU~\cite{papineni2002bleu}& 2002 & General & Match  &  \CheckmarkBold  & \XSolidBrush & \XSolidBrush & - & \XSolidBrush & \CheckmarkBold &\\
& Qiu et al. \cite{6748477} & 2013 & Domain-specific & Match & \CheckmarkBold & \CheckmarkBold  & \XSolidBrush & -  & \XSolidBrush & \XSolidBrush &\\
& NINHS~\cite{6933505} & 2014 & Domain-specific & Match-Embedding  & \CheckmarkBold & \makecell{\CheckmarkBold \\ (exclude param \\ lists \& quantity)}  & \makecell{\CheckmarkBold \\ (only name) } & wordnet & \makecell{\CheckmarkBold \\ (only relationship)} & \XSolidBrush & \\
& Nikiforova et al. \cite{nikiforova2015approach}& 2015 & Domain-specific & Match & \XSolidBrush & \CheckmarkBold  & \XSolidBrush  & - & \CheckmarkBold & \XSolidBrush &\\
& BertScore~\cite{Zhang*2020BERTScore:}& 2020 & General & Embedding & \CheckmarkBold  & \XSolidBrush  & \CheckmarkBold & Bert  & \XSolidBrush & \CheckmarkBold &\\
& Yuan et al.\cite{yuan2020structural}& 2020 & Domain-specific & Match & \CheckmarkBold  & \CheckmarkBold & \XSolidBrush  & - & \XSolidBrush & \XSolidBrush &\\
& CodeBERTScore~\cite{zhou-etal-2023-codebertscore}& 2023 & Domain-specific & Embedding & \CheckmarkBold  & \XSolidBrush  & \CheckmarkBold & CodeBERT  & \XSolidBrush & \CheckmarkBold &\\
\midrule
& \textit{CLUE} & 2026 & Domain-specific & Match-Embedding & \CheckmarkBold  & \makecell{\CheckmarkBold \\ (all)} & \makecell{\CheckmarkBold \\ (all)}  & CodeBERT & \makecell{\CheckmarkBold \\ (all)} & \CheckmarkBold &\\ 
\bottomrule
\end{tabular}
}
}
\begin{flushleft}
\small
  \textbf{CStS}=Consider Structural Similarity? \textbf{CSmS}=Consider Semantic Similarity? \textbf{SCM}=Semantic Computation Model. \textbf{CEMCS}=Can Evaluate Model Component Similarity? 
\end{flushleft}
\vspace{-6mm}
\end{table}

\subsection{Existing Evaluation Methods for OOD Generation}

Unlike code generation, which can be quantitatively evaluated using test cases with metrics such as Pass@k~\cite{chen2021evaluatinglargelanguagemodels} metric, the evaluation of OOD mainly compares reference designs with candidate designs. Existing evaluation methods fall into three categories: (1) N-gram-based text evaluation methods (e.g., BLEU~\cite{papineni2002bleu}, ROUGE~\cite{lin-2004-rouge}), which fail to capture the topological structure and semantic features of class diagrams; (2) Semantic embedding-based evaluation methods (e.g., BertScore~\cite{Zhang*2020BERTScore:}, CodeBERTScore~\cite{zhou-etal-2023-codebertscore}), which can extract overall semantic information but lack analysis of fine-grained local semantics; (3) Domain-specific methods for class diagram design (e.g., ~\cite{6748477,nikiforova2015approach,yuan2020structural,6933505}), which typically leverage topological structure but are limited in semantic handling. For instance, NINHS~\cite{6933505} considers name semantics but overlooks method parameters, relationship quantities, and type semantics. These methods also lack public implementations, hindering reproducibility and widespread adoption. Table~\ref{tab:metric_survey} compares CLUE with existing methods across structural coverage, semantic coverage, evaluation granularity, and reproducibility. Our proposed CLUE evaluation metric accounts for the complete structural and semantic information of class diagrams (including classes, methods, parameters, relationships, and their quantities), overcoming the limitations mentioned above. Additionally, CLUE can assess both global and local similarities, such as class elements, attributes, methods, and relationships, offering greater applicability. The implementation of CLUE and the evaluation dataset are publicly available, providing a standardized evaluation metric for OOD generation techniques.

\subsection{Existing Empirical Studies for OOD Generation}

In recent years, researchers have actively explored the potential of LLMs in automated modeling, such as generating UML class diagrams. Chen et al. \cite{10344012} manually compared GPT-3.5 and GPT-4 on 10 cases across classes, attributes, and relationships; Cámara et al. \cite{camara2023assessment} examined ChatGPT’s performance on UML diagrams and OCL constraints using 8 cases; De Bari et al. \cite{10.1145/3674805.3690741} proposed a multidimensional (syntax, semantics, practicality, correctness) evaluation framework comparing GPT-4 with human designs; and Calamo et al. \cite{10.1007/978-3-031-95397-2_13}, which introduced automated evaluation and analyzed the impact of model scale and task complexity. These studies have deepened understanding of LLMs in object-oriented design and provided valuable evaluation paradigms. However, current research has notable limitations: (1) small-scale experiments (typically $\leq 10$ cases) and narrow model coverage restrict generalizability; (2) reliance on single-expert designs or unvalidated automated metrics introduces subjectivity and uncertainty; (3) insufficient analysis of defect types/frequencies and the near-absence of class method evaluation.

This paper systematically evaluates LLMs’ capabilities in generating UML class diagrams from requirements, significantly extending prior work in scale, methodology, and depth. Specifically, we (1) constructed a benchmark of 50 diverse, multi-difficulty object-oriented design tasks; (2) incorporated 940 real undergraduate designs with instructor scores to better reflect authentic educational scenarios and increase the practical relevance of findings; (3) proposed and validated an automated evaluation method grounded in these human data; (4) comprehensively assessed 29 LLMs across classes, attributes, methods, relationships, and overall diagrams; (5) quantitatively characterized syntactic and semantic defect patterns; (6) compared performance across LLM families to guide model selection; and (7) analyzed correlations between task characteristics and generation quality, offering directions for future dataset expansion.

\section{New Benchmarks}

In this section, we introduce OODEval, a new benchmark for evaluating object-oriented design task, along with its human-rated variant, OODEval-Human, and detail their construction process, format, and key characteristics.

\subsection{OODEval: Benchmark for OOD Evaluation }\label{chap:OODEval}

The OODEval benchmark is designed to evaluate the ability of LLMs to generate class diagrams design from natural language requirements. The primary challenge in constructing a high-quality OOD dataset lies in the collection of software requirements. Publicly available software requirements are already scarce, and resources that provide both requirements and the corresponding design class diagrams are even more limited. Existing datasets (e.g., \textit{OOPEval}\cite{wang2024oop} and \textit{RepoClassBench}\cite{deshpande2024classlevelcodegenerationnatural}) automatically generate requirement descriptions from code. However, the absence of a rigorous requirement modeling process often leads to incomplete or inaccurate results, thereby reducing the quality and applicability of these datasets. To address these limitations, we adopted a manual collection strategy drawing from diverse sources:  (1) leveraging GitHub's native support for the Mermaid modeling language, we performed a keyword search using the query ``mermaid classdiagram language:Markdown AND requirement path:*.md'' to identify Markdown files containing keywords of "mermaid", "classdiagram", and "requirement"; (2) surveying literature on requirements-to-design mappings and extracting usable open-source OOD datasets from appendices and repositories; and (3) compiling requirements and designs from archived projects in university OOD courses. To avoid data leakage,  we restricted collection to data created after 2025, as the latest training data cutoff date of the LLMs studied  in this work is 2024 (see Table~\ref{tab:model-info}). Furthermore, we retained only those instances that contain complete requirements together with their corresponding class diagrams. This process ultimately  yielded 50 high-quality instances (13 from GitHub, 20 from academic  literature, and 17 from university courses), each accompanied by high-quality requirements and design specifications.

Nevertheless, the original class diagrams exhibited deficiencies in completeness and consistency. To ensure quality, three researchers with extensive OOD experience independently reviewed the requirement--design mappings and revised or supplemented class diagrams when necessary. Consensus was then reached through group discussion to form unified final designs, thereby reducing subjective bias. On average, each data review required 2 hours per person, amounting to a total of 100 hours per person across the dataset. All final diagrams were implemented in PlantUML to guarantee representational consistency and reproducibility. In addition, the complete dataset was exported into JSON format using the OODEval construction toolkit to facilitate automated evaluation and downstream applications. The final dataset consists of 50 standardized requirement--design pairs that can be directly applied to OOD-related research.

\subsubsection{Benchmark Format.\quad} 
\begin{figure*}[t]
    \vspace{-2mm}
    \centering
    \includegraphics[width=1.0\linewidth]{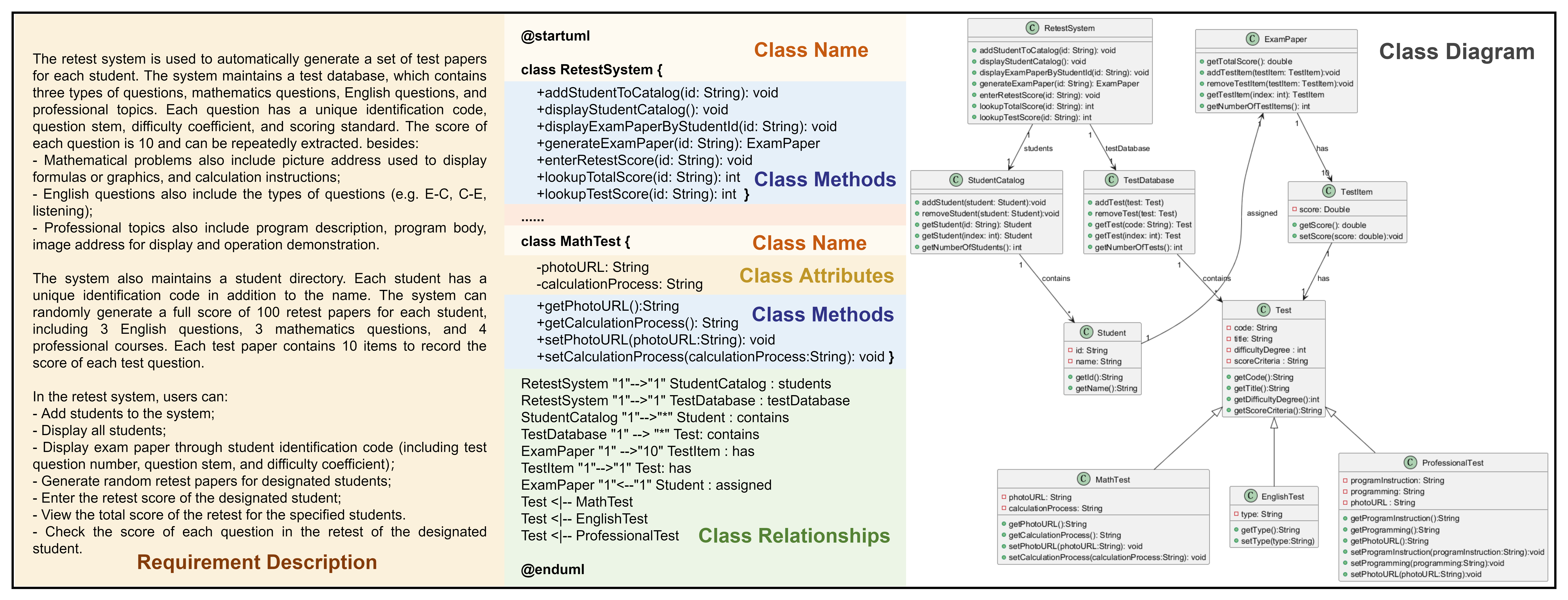}
    \vspace{-6mm}
    \caption{An Example of Input and Output Format in OODEval Benchmark.}
    \label{fig:oodeval_benchmark}
    \vspace{-4mm}
\end{figure*}

Each task in OODEval is structured using our builder toolkit in JSON and PNG formats to facilitate evaluation and visualization. As illustrated in Figure~\ref{fig:oodeval_benchmark}, each task comprises three components: a natural language description of software requirements, a corresponding class diagram in PlantUML code, and a PNG image rendered from the code for intuitive inspection. The PlantUML code is delimited by the \texttt{@startuml} and \texttt{@enduml} tags to enable efficient post-processing. To support fine-grained evaluation of model performance on class diagram elements, we extract metadata from the PlantUML code for each task, including class names, attributes, methods, and relationships. This extraction is enabled by a dedicated PlantUML parser we developed, which extracts the metadata and verifies syntactic correctness. Consequently, OODEval enables both global assessment of overall class diagram quality and detailed evaluation of its individual components, providing a comprehensive analysis of designs generated by large language models.

\subsubsection{Key Characteristics.\quad}
The OODEval benchmark spans 10 primary domains, with approximately 40\% focusing on public service-oriented systems (e.g., food, retail, and transportation), 30\% on management and finance, and the remaining 30\% on specialized fields such as healthcare and computing. This broad coverage facilitates a comprehensive assessment of large language models' (LLMs) capabilities in object-oriented design. Furthermore, OODEval assigns a difficulty rating to each task, derived from key metrics: the number of classes, the number of attributes per class, the number of methods per class, the numbe of relationships, and requirement readability (measured via the Flesch-Kincaid Reading Ease Score~\cite{article_360855}). Tasks with higher counts of classes, attributes, methods, and relationships, alongside lower readability scores, are deemed more challenging. The difficulty rating is computed by normalizing these metrics and calculating a weighted average using cross-entropy weights~\cite{CHEN2021114186}. To examine LLMs' performance across varying complexities, tasks are stratified into simple, moderate, and hard categories based on the 33rd and 67th percentiles of the difficulty ratings. Table~\ref{tab:oodeval-characteristics} presents the mean and standard error of each metric across these subsets and the overall dataset.

\begin{table}[ht]
\vspace{-3mm}
\setlength{\tabcolsep}{7.5pt} 
\caption{Statistics of the OODEval Benchmark Across Difficulty Categories.}
\vspace{-3mm}
\label{tab:oodeval-characteristics}
{\small
\begin{tabular}{@{\extracolsep{\fill}}llccccccll@{\extracolsep{\fill}}}
\toprule
& \textbf{Set} & \textbf{\# Data} & \textbf{\# Class} & \textbf{\# Avg Attr} & \textbf{\# Avg Mtd} & \textbf{\# Relation} & \textbf{Readability} & \textbf{Difficulty} &\\ 
\midrule
& simple & 16 & 6.50$\pm$1.67 & 2.30$\pm$1.05 & 0.69$\pm$0.97 & 7.38$\pm$2.70 & 62.44$\pm$6.72 & $ [0.0,0.2800)$ &\\
& moderate & 17 & 7.18$\pm$3.21 & 2.28$\pm$0.83 & 2.12$\pm$1.36 & 8.88$\pm$4.23 & 50.39$\pm$11.17 & $ [0.2800,0.4583)$ &\\
& hard & 17 & 10.18$\pm$2.63 & 2.55$\pm$0.88 & 3.40$\pm$1.42 & 12.00$\pm$3.43 & 42.57$\pm$11.74 & $[0.4583,1.0)$ &\\ \midrule
& overall & 50 & 7.98$\pm$3.01 & 2.38$\pm$0.91 & 2.10$\pm$1.67 & 9.46$\pm$3.96 & 51.47$\pm$12.89 & $[0.0,1.0)$ &\\
\bottomrule
\end{tabular}
}
\vspace{-3mm}
\end{table}

\subsection{OODEval-Human: Undergraduate OOD Solutions with Instructor Ratings}

\subsubsection{Construction Procedure.\quad}  The OODEval-Human dataset is collected for two primary goals: (1) to evaluate whether LLMs can reach the level of undergraduate students on OOD tasks, and (2) to assess the effectiveness of the CLUE metric, since each solution is given a well-considered and consistent instuctor rating. Specifically, we collected and cleaned over 1K undergraduate submissions from four OOD tasks, including both student solutions and their corresponding scores, which were obtained from course archives. These four OOD tasks are part of the OODEval benchmark, where they are paired with reference design solutions for comparative analysis. All submissions were graded using unified and consistent scoring criteria by multiple instructors.

Since the original submissions are primarily class diagrams in image format, they are unsuitable for automated processing and evaluation. It was therefore necessary to convert these images into PlantUML code format. However, performing this conversion manually would have been extremely time-consuming. Benefiting from the multimodal capabilities of LLMs such as GPT-4o, we first encoded the submitted images into Base64 format and then employed the GPT-4o model to automatically generate the corresponding PlantUML representations from these images. However, subtle discrepancies (such as missing or misused relationships, inconsistent class/attribute/method names, etc.) were observed between the automatically generated PlantUML code and the students’ original class diagrams, which limited the reliability of direct use.

To ensure consistency between the image and PlantUML code, we designed a manual verification process. Specifically, the OODEval builder toolkit was used to render the generated PlantUML code into an image (denoted as Image $B$), which was then compared with the original student submission (denoted as Image $A$). The three human reviewers manually inspected the consistency between Image $A$ and Image $B$ and, when necessary, revised the PlantUML code to guarantee alignment between the rendered diagram and the original submission. After the three human reviewers completed modifications across all submissions, an additional independent reviewer, who did not participate in the initial review, conducted a final verification to avoid any omissions or errors, ensuring complete consistency between all Image $A$ and Image $B$ pairs. In this way, only partial elements needed updating compared to creating all elements from scratch, resulting in an average time of 10 minutes per review and a total of nearly 160 hours across the entire process.

The validated outputs are subsequently exported as standardized dataset entries. Through this combined process of automatic generation and manual verification, we systematically validated all original data and removed invalid cases where Image A was incomplete or of insufficient quality. In total, we obtained OODEval-Human, which consists of 940 high-quality validated data instances, each containing a student solution together with its associated instructor rating.

\subsubsection{Benchmark Format.\quad} The format of the OODEval-Human dataset closely aligns with that of OODEval, with the key distinction being the design solution. While the design solution in OODEval consists of a reference design class diagram, in OODEval-Human, the design solution is an undergraduate submission class diagram. Additionally, OODEval-Human includes supplementary information. Specifically, it retains the original student submissions in image format, along with the ratings assigned by instructors. This design allows OODEval-Human to support the performance evaluation for automated evaluation metrics (e.g. the CLUE metrics in this paper) on requirement-to-design generation tasks, while also serving as a valuable dataset for analyzing the differences and performance gaps between LLMs and human designers in OOD tasks.

\subsubsection{Key Characteristics.\quad} Table~\ref{tab:oodeval-human-characteristics} presents the statistics of the OODEval-Human dataset, which consists of four undergraduate OOD tasks, each associated with OODEval by its Task ID number. Each task includes over 200 student submissions, totaling 940 submissions, with individual scores ranging from 32 to 100. The average scores for the tasks range from 79.3 to 82.8, with an overall average of 80.62. The CLUE metric serves as an indicator for quantifying the similarity between two designs; thus, it is independent of requirement diversity. Consequently, the distinct solutions submitted by each student ensure a sufficient sample size and diverse score distribution, thereby enabling robust evaluation of the CLUE metric's performance, even though OODEval-Human comprises four tasks. Furthermore, the difficulty levels of these four tasks in OODEval are moderate to hard, making them well-suited for benchmarking the OOD capabilities of LLMs against human performance on more challenging tasks.
\begin{table}[ht]
\vspace{-3mm}
\setlength{\tabcolsep}{21pt} 
\caption{Statistics of the OODEval-Human Dataset.}
\vspace{-2mm}
\label{tab:oodeval-human-characteristics}
{\small
\begin{tabular}{@{\extracolsep{\fill}}lccccll@{\extracolsep{\fill}}}
\toprule
 & \textbf{Task ID}  & \textbf{\# Data} & \textbf{Score Range} & \textbf{Avg Score} & \textbf{Difficulty} &\\ \midrule
 & 47  & 236 & [32,100] & 79.91 & 0.54(hard) &\\
 & 48  & 240 & [54,96] & 79.30   & 0.41(moderate) &\\
 & 49  & 214 & [50,96] & 82.82  & 0.82(hard) &\\
 & 50  & 250 & [61,93] & 80.68  & 0.47(hard) &\\ \midrule
 & Total & 940 & [32,100] & 80.62 & 0.56(hard) & \\ 
 \bottomrule
\end{tabular}
}
\vspace{-3mm}
\end{table}

\section{CLUE: Class Likeness Unified Evaluation Metric}

This section introduces a novel evaluation metric for assessing the correctness of class diagram designs. First, we present the preliminary definitions required for the formulation of CLUE. We then describe the computation of the CLUE metric, followed by a hyperparameter optimization procedure to determine the optimal weights. To evaluate the effectiveness of CLUE, we adopt commonly used metric evaluation practices reported in prior work~\cite{10.1016/j.jss.2023.111741, zhou-etal-2023-codebertscore, xu2024human}, which assess the correlation between automatic metrics and human preferences; a higher correlation indicates a better metric. In this study, we compare CLUE scores with instructor ratings on the OODEval-Human dataset and observe significantly higher correlation coefficients than those achieved by other metrics.

\subsection{Preliminaries.\quad} 

The CLUE metric is specifically devised to measure the degree of structural and semantic similarity between a candidate class diagram and a corresponding reference class diagram. Accordingly, we begin by formally defining the concept of a Class Diagram Model, as articulated in Definition~\ref{def:class_diagram_model}.

\begin{definition}[Class Diagram Model]\label{def:class_diagram_model}
A class diagram model is formally defined as a tuple \( \mathcal{D} = (C, R),\) where:
\begin{itemize}
    \item \( C = \{c_1, c_2, \dots, c_n\} \) is a finite set of classes, and each class \(c = (name, A, M),\) consists of a class name,a finite set of attributes \( A \), and a finite set of methods \( M \). An attribute is a pair \( a = (\textit{name}, \textit{type}) \), where both elements are strings. A method is a triple \( m = (\textit{name}, \textit{returnType}, P) \), where \( \textit{name} \) and \( \textit{returnType} \) are strings, and \( P = \{p_1, \dots, p_k\} \) is a finite set of parameters, each parameter being a pair \( p = (\textit{name}, \textit{type}) \) of strings.

    \item \( R = \{r_1, r_2, \dots, r_s\} \) is a finite set of relationships, where each relationship is defined as \( r = (\textit{type}, \textit{begin}, \textit{end}, \textit{label}) \). Here, \( \textit{type} \in \mathbb{T} = \{\text{AS}, \text{AG}, \text{CO}, \text{DE}, \text{GE}, \text{RE}\} \) denotes the relationship type (association, aggregation, composition, dependency, generalization, and realization). \( \textit{begin}, \textit{end} \in \{1,\dots,|C|\} \) denote the indices of the source and target classes in \(C\), respectively. And \( \textit{label} = (\textit{from}, \textit{to}) \) specifies the multiplicity constraints at both ends.
\end{itemize}
\end{definition}

Computing the similarity between two Class Diagram Models poses substantial challenges, primarily arising from the fact that semantically equivalent model elements frequently exhibit differing lexical names. For example, the classes \textit{Teacher} and \textit{Tutor} may denote essentially the same conceptual entity despite their distinct labels, which is a common phenomenon in software design practice. When the elements under comparison consist solely of strings (e.g., class names, attribute names, or operation signatures), semantic similarity can be effectively captured by computing the cosine similarity between their corresponding embedding vectors, as formalized in Definition~\ref{def:string_semantic_similarity}. In this work, we adopt the CodeBERT model \cite{feng-etal-2020-codebert} as the embedding function to generate high-quality contextual embeddings for input strings $  s_1  $ and $  s_2  $. This choice is motivated by CodeBERT's superior performance in capturing underlying semantics, as demonstrated through an ablation study evaluating various underlying semantic models (see Section~\ref{sec:ablation_study_on_semantic}).

\begin{definition}[String Semantic Similarity]\label{def:string_semantic_similarity}
Let $\mathbf{e}: \mathcal{S} \to \mathbb{R}^d$ be an embedding function that maps strings to $d$-dimensional real vectors, where $\mathcal{S}$ denotes the space of strings. The semantic similarity $\sigma(s_1, s_2) \in [0,1]$ between strings $s_1, s_2 \in \mathcal{S}$ is given by normalized cosine similarity between their embedding vectors:
\begin{equation}
\sigma(s_1, s_2) = 0.5 \cdot ( \frac{\mathbf{e}(s_1)^\top \mathbf{e}(s_2)}{\|\mathbf{e}(s_1)\|_2 \|\mathbf{e}(s_2)\|_2} + 1)
\end{equation}
where $\cdot^\top$ denotes the transpose, and $\|\cdot\|_2$ is the Euclidean norm.
\end{definition}

When comparing two sets of model elements (e.g., the attribute sets of a reference class and a candidate class), the task becomes significantly more challenging. Simple exact name matching is clearly insufficient, as semantically equivalent attributes may carry different names. To address this, we first compute the pairwise similarities between every element of the reference set and every element of the candidate set, thereby constructing the Element Set Similarity Matrix as formally defined in Definition~\ref{def:element_set_similarity}. Subsequently, we identify an optimal matching $\pi$ between the two sets that maximizes the sum of the similarities of the matched pairs (see Definition~\ref{def:optimal_matching_similarity}).

\begin{definition}[Element Set Similarity Matrix]\label{def:element_set_similarity}
Let \( A = \{a_1, \dots, a_n\} \) and \( B = \{b_1, \dots, b_m\} \) be two element sets. Each element \( a_i \in A \) and \( b_j \in B \) is represented by a \( d \)-tuple \((c_1, \dots, c_d).\) Let \( \boldsymbol{\mathrm{sim}} = (\mathrm{sim}_1, \dots, \mathrm{sim}_d) \)
denote a collection of similarity functions, and let \( \mathbf{w} = (w_1, \dots, w_d) \) be a weight vector satisfying \( w_k \ge 0 \) and \( \sum_{k=1}^d w_k = 1 \). The element set similarity matrix between \( A \) and \( B \) is defined as \( ES \in [0,1]^{n \times m} \), where each entry \(ES(i,j) \) is computed as:
\begin{equation}\label{eq:3}
ES(i,j)
=
\sum_{k=1}^{d}
w_k \cdot
\mathrm{sim}_k\bigl(c_k(a_i), c_k(b_j)\bigr),
\quad
i = 1,\dots,n,\; j = 1,\dots,m .
\end{equation}
where each similarity function \( \mathrm{sim}_k(\cdot,\cdot) \) maps two strings (Definition~\ref{def:string_semantic_similarity}), two elements (Definition~\ref{def:multiplicity_similarity}), or two element sets (Definition~\ref{def:optimal_matching_similarity}) to a value in \( [0,1] \), depending on the structural type of the \( k \)-th component.
\end{definition}

\begin{definition}[Optimal Matching Similarity]\label{def:optimal_matching_similarity}
Let $R = \{r_1, r_2, \dots, r_n\}$ be the reference set and $C = \{c_1, c_2, \dots, c_m\}$ be the candidate set, where $n>0$ and $m>0$. Let $ES \in [0,1]^{n \times m}$ be the similarity matrix between $R$ and $C$. Each entry $ES(i,j)\in[0,1]$ represents the similarity between the $i$-th element of $R$ and the $j$-th element of $C$(see Definition~\ref{def:element_set_similarity}). The optimal matching similarity $OptMatch(R,C)\in[0,1]$ is then defined as:

\begin{equation}
OptMatch(R,C) = \begin{cases} 
\frac{1}{n}\max\limits_{\pi} \left\{ \sum_{i=1}^n ES(i,\pi(i)) \right\}, & \text{if } m\geq n>0 \\
\frac{1}{n}\max\limits_{\pi'} \left\{ \sum_{j=1}^m ES(\pi'(j),j) \right\}, & \text{if } n>m>0
\end{cases}
\label{formula:optimal_matching}
\end{equation}
where \( \pi(·) \) is an injective function from $R$ to $C$, and \(\pi'(·)\) is an injective function from $C$ to $R$. 
\end{definition}

Note that, irrespective of whether \( n > m \) or \( n \leq m \), the optimal matching similarity is normalized by \( n \), where \( n \) denotes the cardinality of the element set in the reference model. This design choice fundamentally distinguishes CLUE from other similarity metrics and underscores the mandatory presence of a designated reference model. In the CLUE algorithm, similarity computation is inherently anchored to the reference model: missing elements in the candidate (i.e., \( n > m \)) incur a penalty, whereas extraneous elements in the candidate (i.e., \( n \leq m \)) are disregarded. In this paper, we select the Hungarian algorithm\cite{9548600} to implement the optimal matching due to its high speed with \(O(n^3)\) complexity and its guarantee of a global optimum. It is particularly suitable when $n$ is not large, and in this paper $n<20$.

In class relationship designing, multiplicity constraints between two classes are typically denoted by numeric or symbolic labels at each end of the relationship. However, the string semantic similarity measure introduced earlier in Definition~\ref{def:string_semantic_similarity} fails to adequately capture the similarity of these constraints. To address this limitation, we define relationship multiplicity similarity for any pair of class relationships, as a domain-informed heuristic grounded in UML semantics, as shown in Definition~\ref{def:multiplicity_similarity}.

\begin{definition}[Relationship Multiplicity Similarity]\label{def:multiplicity_similarity}
Let $\mathbb{T}_1 = \{\text{AS}, \text{AG}, \text{CO}\}$ and $\mathbb{T}_2 = \{\text{DE}, \text{GE}, \text{RE}\}$, where $\mathbb{T}_1 \cup \mathbb{T}_2 = \mathbb{T}$. Given two relationship multiplicity labels \(\ell_1=(from_1, to_1)\) and \(\ell_2=(from_2, to_2)\), and their corresponding relationship types \(t_1\) and \(t_2\), the relationship multiplicity similarity is defined as:
\begin{equation}
\operatorname{sim}_{rm}(\ell_1, \ell_2 \mid t_1, t_2) =
\begin{cases}
0.5 \cdot \bigl(\operatorname{match}(from_1, from_2) + \operatorname{match}(to_1, to_2)\bigr),
& \text{if } t_1, t_2 \in \mathbb{T}_1, \\[1mm]
1.0, & \text{if } t_1, t_2 \in \mathbb{T}_2, \\[1mm]
0.0, & \text{otherwise}.
\end{cases}
\end{equation}
where \( \operatorname{match}(q_1, q_2) = 1 \) if \( q_1 = q_2 \) or both \( q_1 \) and \( q_2 \) match ``*'', ``many'', ``much'', or ``multi'' via regular expression, and 0 otherwise.
\end{definition}
\vspace{-3mm}

\subsection{CLUE Metrics.\quad} \label{subsubsec:clue_metrics}

CLUE represents a comprehensive suite of metrics designed to evaluate class diagrams at multiple granularities (see Definition~\ref{def:clue_metric}): the primary $\textbf{\textit{clue}}$ metric assesses overall diagram similarity; $\textbf{\textit{clue-class}}$ focuses on individual class similarity; $\textbf{\textit{clue-attribute}}$ evaluates the similarity of class attributes; $\textbf{\textit{clue-method}}$ measures the similarity of class methods; and $\textbf{\textit{clue-relation}}$ quantifies the similarity of inter-class relationships. To avoid nomenclature ambiguity, we use uppercase CLUE to denote the entire metric suite, while lowercase clue specifically refers to the overarching metric for class diagram similarity assessment. Compared to existing methods, CLUE distinctively integrates both structural aspects of class diagrams and enriched semantic representations, while incorporating fine-grained semantic computations within the structural matching phase, thereby enabling a more accurate and fine-grained evaluation. 

\begin{definition}[Class Likeness Unified Evaluation Metric, \( \text{CLUE} \)]\label{def:clue_metric}
Given a reference class diagram model \(\mathcal{D}_1=(C_1,R_1)\) and a candidate model \(\mathcal{D}_2=(C_2,R_2)\), where \(C_1 = (c^1_1,\dots, c^1_n)\) and \(C_2 = (c^2_1,\dots, c^2_m)\), each class \(c^1_i \in C_1\) is defined as \(c^1_i=(name^1_i,A^1_i,M^1_i)\) and each class \(c^2_j \in C_2\) is defined as \(c^2_j=(name^2_j,A^2_j,M^2_j)\). The CLUE metric suite consists of five interrelated similarity metrics defined as:

\begin{equation}
clue(\mathcal{D}_1,\mathcal{D}_2)=w_e\cdot clue\mbox{-}class(\mathcal{D}_1,\mathcal{D}_2)+w_r\cdot clue\mbox{-}relation(\mathcal{D}_1,\mathcal{D}_2)
\end{equation}
\begin{equation}
\textit{clue-class}(\mathcal{D}_1, \mathcal{D}_2) = OptMatch(C_1,C_2)
\end{equation}
\begin{equation}
\textit{clue-relation}(\mathcal{D}_1, \mathcal{D}_2) = OptMatch(R_1, R_2)
\end{equation}
\begin{equation}
clue\mbox{-}attribute(\mathcal{D}_1,\mathcal{D}_2) = \begin{cases} 
\frac{1}{n} \sum_{i=1}^{n} OptMatch(A^1_i, A^2_{\pi(i)}), & \text{if } m\geq n>0 \\
\frac{1}{n} \sum_{j=1}^{m} OptMatch(A^1_{\pi'(j)}, A^2_j), & \text{if } n>m>0
\end{cases}
\end{equation}
\begin{equation}
clue\mbox{-}method(\mathcal{D}_1,\mathcal{D}_2) = \begin{cases} 
\frac{1}{n} \sum_{i=1}^{n} OptMatch(M^1_i, M^2_{\pi(i)}), & \text{if } m\geq n>0 \\
\frac{1}{n} \sum_{j=1}^{m} OptMatch(M^1_{\pi'(j)}, M^2_j), & \text{if } n>m>0
\end{cases}
\end{equation}
where \( \pi(\cdot) \) and \( \pi'(\cdot) \) denote the injective mappings induced by the optimal class matching between \(C_1\) and \(C_2\), as defined in Definition~\ref{def:optimal_matching_similarity}.
\end{definition}

\begin{figure*}[t]
    \vspace{-2mm}
    \centering
    \includegraphics[width=1.0\linewidth]{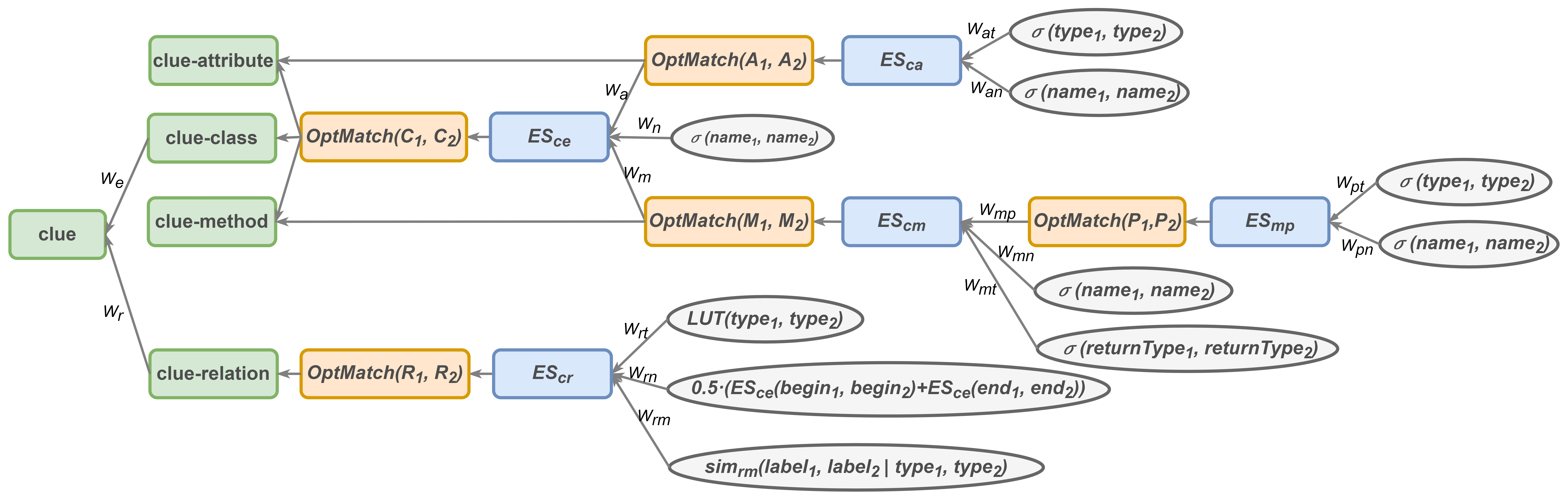}
    \vspace{-6mm}
    \caption{Computational Dependency Graph of CLUE Metrics.}
    \label{fig:computing_dependency}
    \vspace{-4mm}
\end{figure*}

Figure~\ref{fig:computing_dependency} illustrates the computation dependency graph for the five aforementioned CLUE metrics. The computation of CLUE follows a multi-layer recursive process with layer-wise weighting. Starting from the root node \textbf{clue}, the recursion proceeds with a maximum depth of four layers. Each recursive edge in the graph is associated with a tunable weight parameter, resulting in a total of 15 trainable weights across the entire dependency structure. In the figure, \textbf{green nodes} represent the five core metrics of CLUE (clue-class, clue-attribute, clue-method, clue-relation, and the overall clue); \textbf{blue nodes} denote the similarity matrices that require optimal matching, as formally defined in Table~\ref{tab:element_similarity_matrix}; \textbf{orange nodes} represent the optimal matching similarity on $ES$ matrix between two sets; and \textbf{gray nodes} correspond to the leaf nodes of the recursive computation, each of which implements an atomic similarity function serving as the base case of the CLUE metric.

As illustrated in Figure~\ref{fig:computing_dependency}, CLUE requires five types of similarity matrices, namely \( ES_{ce} \), \( ES_{ca} \), \( ES_{cm} \), \( ES_{mp} \), and \( ES_{cr} \), which are computed over class sets, relationship sets, attribute sets, method sets, and parameter sets, respectively. All these similarity matrices are defined as specific instantiations of the Element Set Similarity Matrix (Definition~\ref{def:element_set_similarity}) by instantiating the four aspects: (1) the element sets \( A \) and \( B \); (2) the internal representation $d\mbox{-}tuple$ of each element in \( A \) and \( B \); (3) the corresponding weight vector \( \mathbf{w} \); and (4) the corresponding similarity functions. Table~\ref{tab:element_similarity_matrix} summarizes the instantiation details for these five types of similarity matrices.

\begin{table}[ht]
\vspace{-3mm}
\setlength{\tabcolsep}{10.5pt}
\caption{Instantiation for Element Set Similarity Matrices.}
\vspace{-2mm}
\label{tab:element_similarity_matrix}
{\small
\begin{tabular}{@{\extracolsep{\fill}} lllllll @{\extracolsep{\fill}}}
\toprule
 & \textbf{\textit{ES}}   & \textbf{Element Sets} & \textbf{\textit{d}-tuple} & \textbf{weight vector} & \textbf{similarity function} &\\ 
 \midrule
 & $ES_{ce}$  & class sets & $(name,A,M)$ & $(w_n,w_a,w_m)$ & $(\sigma,OptMatch,OptMatch)$ &\\
 & $ES_{ca}$  & attribute sets & $(name,type)$ & $(w_{an},w_{at})$  & $(\sigma,\sigma)$ &\\
 & $ES_{cm}$  & method sets & $(name,returnType,P)$ & $(w_{mn},w_{mt},w_{mp})$ & $(\sigma,\sigma, OptMatch)$ &\\ 
 & $ES_{mp}$  & parameter sets & $(name,type)$ & $(w_{pn},w_{pt})$ & $(\sigma,\sigma)$ &\\ 
 & $ES_{cr}$  & relationship sets & $(type,begin,end,label)$ & $(w_{rt},w_{rn}/2,w_{rn}/2,w_{rm})$ & $(LUT,ES_{ce},ES_{ce},\operatorname{sim}_{rm})$ &\\
 \bottomrule
\addlinespace[1.5pt]
\multicolumn{5}{@{}l}{\footnotesize all of the notions in the table come from Definition~\ref{def:class_diagram_model} to \ref{def:class_relationship_similarity}.}
\end{tabular}
}
\vspace{-3mm}
\end{table}

The similarity matrices $  ES_{ce}  $, $  ES_{ca}  $, $  ES_{cm}  $, and $  ES_{mp}  $ adhere to a unified computational framework. Specifically, the string similarity function $  \sigma(\cdot, \cdot)  $ is employed for pairwise comparisons of string-valued elements, whereas the optimal matching similarity $  \text{OptMatch}(\cdot, \cdot)  $ is utilized for comparing sets of elements. The computation proceeds in a bottom-up, recursive pattern: $  ES_{mp}  $ is calculated first, followed by $  ES_{cm}  $ and $  ES_{ca}  $, with the results ultimately aggregated into $  ES_{ce}  $. This hierarchical design guarantees local optimality at each level, which is then propagated to subsequent higher-level aggregations. In contrast, the relationship similarity matrix $  ES_{cr}  $ adopts a distinct formulation. This stems from the fact that relationship similarity is contingent not only on intrinsic attributes (e.g., relationship type and multiplicity) but also on the similarity of the classes connected at both endpoints. Consequently, the viability of a relationship match is profoundly influenced by the semantic alignment of its source and target classes. To precisely delineate the associated notation in $  ES_{cr}  $ and , we formally define the relationship similarity matrix in Definition~\ref{def:class_relationship_similarity}.

\begin{definition}[Relationship Similarity Matrix]\label{def:class_relationship_similarity}
Let \( R_1 = \{r^1_1, \dots, r^1_p\} \) and \( R_2 = \{r^2_1, \dots, r^2_q\} \) be the reference and candidate relationship sets, respectively. Each relationship \( r = (\textit{type}, \textit{begin}, \textit{end}, \textit{label}) \) follows the definition in the class diagram model (Definition~\ref{def:class_diagram_model}). The relationship similarity matrix is defined as \( ES_{cr} \in [0,1]^{p \times q} \), where each entry \( ES_{cr}(i,j) \) is computed as:
\begin{equation}\label{eq:8}
ES_{cr}(i,j) = w_{rt} \cdot \mathrm{LUT}(type^1_i, type^2_j) + \frac{w_{rn}}{2} \cdot \left( ES_{ce}(begin^1_i, begin^2_j) + ES_{ce}(end^1_i, end^2_j) \right) + w_{rm}  \cdot \mathrm{sim}_{rm}(label^1_i, label^2_j) 
\end{equation}
for \( i = 1,\dots,p \) and \( j = 1,\dots,q \), where \( w_{rt}, w_{rm}, w_{rn} \ge 0 \) and
\( w_{rt} + w_{rm} + w_{rn} = 1 \). Here, \( \mathrm{LUT}: \mathbb{T} \times \mathbb{T} \rightarrow [0,1] \) denotes the relationship type similarity defined by a lookup table following the configuration in~\cite{6933505}. The function \( \mathrm{sim}_{rm}(\cdot,\cdot)\) measures the similarity between relationship multiplicity labels (Definition~\ref{def:multiplicity_similarity}). The term \( ES_{ce}(i,j) \) refers to the similarity between the \(i\)-th reference class and the \(j\)-th candidate class in the class similarity matrix.
\end{definition}

\subsection{Hyperparameter Optimization}\label{sec:hyperparameter}

Before applying CLUE, an unresolved question was how to determine, within the 15-dimensional parameter space, a set of parameters that maximizes the correlation between CLUE scores and human ratings. Since the parameters belong to the real number domain, it is not possible to find the best configuration manually. To systematically address this issue, we formulate the determination of the optimal weight configuration as a constrained optimization problem, as presented in Equation~\ref{eq:optimal_problem}.

\begin{equation}\label{eq:optimal_problem}
\left\{
\begin{aligned}
\boldsymbol{x}^* =& \arg\max_{\boldsymbol{x} \in \mathcal{X}} f(\boldsymbol{x}) \\
\text{s.t.} \quad &\boldsymbol{g}(\boldsymbol{x}) \leq \boldsymbol{0}
\end{aligned}
\right.
\end{equation}

\begin{equation}\label{eq:objective_function}
f(\boldsymbol{x}) = \sum_{m \in \mathcal{M}} \rho \left( C_m(\boldsymbol{x}), \boldsymbol{y}_{\text{human}} \right) 
\end{equation}

\begin{equation}\label{eq:g_x}
\left\{\quad
\begin{array}{@{} r@{\,}l @{\quad} r@{\,}l @{\quad} r@{\,}l @{}}
g_1(\boldsymbol{x})  &{}= -a\leq 0           & g_7(\boldsymbol{x})  &{}= -d \leq 0          & g_{13}(\boldsymbol{x}) &{}= -g \leq 0         \\[3pt]
g_2(\boldsymbol{x})  &{}= a-1\leq 0          & g_8(\boldsymbol{x})  &{}= d-1\leq 0          & g_{14}(\boldsymbol{x}) &{}= g-1 \leq 0        \\[3pt]
g_3(\boldsymbol{x})  &{}= -b\leq 0           & g_9(\boldsymbol{x})  &{}= -e\leq 0           & g_{15}(\boldsymbol{x}) &{}= -h\leq 0          \\[3pt]
g_4(\boldsymbol{x})  &{}= b-1\leq 0          & g_{10}(\boldsymbol{x})&{}= e-1\leq 0          & g_{16}(\boldsymbol{x}) &{}= h-1\leq 0         \\[3pt]
g_5(\boldsymbol{x})  &{}= b+c-1\leq 0        & g_{11}(\boldsymbol{x})&{}= e+f-1\leq 0        & g_{17}(\boldsymbol{x}) &{}= h+k-1\leq 0       \\[3pt]
g_6(\boldsymbol{x})  &{}= -c \leq 0          & g_{12}(\boldsymbol{x})&{}= -f\leq 0           & g_{18}(\boldsymbol{x}) &{}= -k \leq 0         
\end{array}
\quad\right. 
\end{equation}

Specifically, $\boldsymbol{x} = (a, b, c, d, e, f, g, h, k) \in \mathbb{R}^9$ represents the 9-dimensional decision variable vector parameterizing the weights $w_e$, $w_n$, $w_a$, $w_{at}$, $w_{mp}$, $w_{mn}$, $w_{pt}$, $w_{rt}$, and $w_{rm}$, respectively. The remaining six weights are derived via normalization constraints, partitioning the weights into groups that sum to 1 (e.g., $w_r = 1 - w_e$ for pairwise groups and $w_m = 1 - w_n - w_a$ for three-weight groups), thus avoiding explicit optimization for complementary weights. The feasible region $\mathcal{X} \subseteq \mathbb{R}^9$ is defined by the 18-dimensional constraint vector $\boldsymbol{g}(\boldsymbol{x})$, comprising inequalities $g_1$ through $g_{18}$ as detailed in Equation~\ref{eq:g_x}. The objective function $f(\boldsymbol{x})$ is defined in Equation~\ref{eq:objective_function}. $\mathcal{M}$ denote the set of evaluation metric types (i.e., clue, clue-attribute, clue-method, clue-class, clue-relation). The mapping $C: \mathbb{R}^9 \times \mathcal{M} \to \mathbb{R}^n$ yields $C_m(\boldsymbol{x})$, the CLUE score vector for metric type $m \in \mathcal{M}$ under configuration $\boldsymbol{x}$. The vector $\boldsymbol{y}_{\text{human}} \in \mathbb{R}^n$ represents the corresponding human-rated scores on the same samples. The Pearson correlation coefficient $\rho(\boldsymbol{u}, \boldsymbol{v}): \mathbb{R}^n \times \mathbb{R}^n \to [-1, 1]$ quantifies the linear correlation between vectors $\boldsymbol{u}$ and $\boldsymbol{v}$. 

\begin{table}[t]
\vspace{-2mm}
\setlength{\tabcolsep}{16.5pt}
\caption{Correlation Coefficient and Statistical Significance On Train and Test Sets.}
\vspace{-2mm}
\label{tab:correlation_on_train_test}
{\small
\begin{tabular}{@{\extracolsep{\fill}} llccccc @{\extracolsep{\fill}}}
\toprule
  & & \multicolumn{2}{c}{\textbf{Train Set (751 items)}} & \multicolumn{2}{c}{\textbf{Test Set (189 items)}} &\\
 \cmidrule(lr){3-4} \cmidrule(lr){5-6}
  & \textbf{Metric} & \textbf{Pearson($\rho$)} & \textbf{Significance($p$)} & \textbf{Pearson($\rho$)} & \textbf{Significance($p$)} &\\ 
\midrule
 & clue-class     & 0.579988 & 1.016773e-68 & 0.605792 & 2.579184e-20 &\\
 & clue-attribute & 0.370033 & 8.770272e-26 & 0.400179 & 1.161585e-08 &\\
 & clue-method    & 0.392214 & 5.017631e-29 & 0.381030 & 6.338065e-08 &\\
 & clue-relation  & 0.432177 & 1.584394e-35 & 0.505889 & 1.125055e-13 &\\
 & clue           & 0.589745 & 1.508446e-71 & 0.617803 & 2.814116e-21 &\\
\bottomrule
\end{tabular}
}
\end{table}

\begin{table}[t]
\vspace{-2mm}
\setlength{\tabcolsep}{15pt}
\caption{Correlations of Baselines and CLUE Metrics with Human Ratings.}
\vspace{-2mm}
\label{tab:correlation_coefficient}
{\small
\begin{tabular}{@{\extracolsep{\fill}} llccccc @{\extracolsep{\fill}}}
\toprule
 & \textbf{Metric} & \textbf{Pearson($\rho$)} & \textbf{Significance($p$)} & \textbf{Spearman($\rho$)} & \textbf{Siginificance($p$)} & \\ 
 \midrule
 & BLEU           & 0.388496  & 3.151621e-35 & 0.323807 & 2.190154e-24 &\\
 & CodeBERTScore  & 0.166168  & 2.999022e-07 & 0.160961 & 7.020592e-07 &\\
 & NINHS          & 0.074554  & 2.225904e-02 & 0.034780 & 2.864958e-01 &\\
\midrule
 & clue-class     & 0.584867  & 8.936066e-65 & 0.408789 & 3.613194e-39 &\\
 & clue-attribute & 0.375269  & 4.315892e-33 & 0.277789 & 4.068201e-18 &\\ 
 & clue-method    & 0.388763  & 5.738255e-34 & 0.330894 & 1.874246e-25 &\\ 
 & clue-relation  & 0.447774  & 8.031337e-46 & 0.421780 & 7.736122e-42 &\\ 
 & clue           & 0.594927* & 4.561636e-91\dag & 0.475737* & 2.954398e-54\dag &\\ 
\bottomrule
\addlinespace[1.5pt]
\multicolumn{5}{@{}l}{\footnotesize *~denotes the maximum value; {\dag}~denotes the minimum value.}
\end{tabular}
}
\end{table}

\begin{table}[t]
\setlength{\tabcolsep}{13pt}
\caption{Ablation Study on the Influence of the Underlying Semantic Model.}
\vspace{-2mm}
\label{tab:ablation_study}
{\small
\centering
\begin{tabular}{@{\extracolsep{\fill}} llcccccc @{\extracolsep{\fill}}}
\toprule
& \textbf{Models} & \textbf{clue-class} & \textbf{clue-attribute} & \textbf{clue-method} & \textbf{clue-relation} & \textbf{clue} &\\ 
\midrule
 & MiniLM    & 0.398578 & 0.377008 & 0.317816 & 0.459131 & 0.435976 & \\
 & BGE       & 0.467757 & 0.400091* & 0.336586 & 0.462346* & 0.496893 & \\
 & BERT      & 0.573388 & 0.389749 & 0.341215 & 0.460218 & 0.589596 & \\
 \midrule
 & CodeBERT  & 0.584867* & 0.375269 & 0.388763* & 0.447774 & 0.594927* & \\
\bottomrule
\addlinespace[1.5pt]
\multicolumn{6}{@{}l}{\footnotesize *~denotes the maximum value}
\end{tabular}
}
\end{table}

The computation procedure of the proposed CLUE metric is highly complex and does not allow explicit differentiation, which makes the objective function \(f(x)\) essentially a black-box function. In addition, each iteration of \(f(x)\) requires computing the Pearson correlation between CLUE scores and human ratings over the entire training dataset, resulting in considerable computational cost. Bayesian Optimization \cite{10.1007/s11334-023-00540-3} is an efficient optimization technique designed for such black-box functions. It constructs a surrogate model, such as a Gaussian Process, to approximate the true objective function and guide the search for the global optimum. Bayesian Optimization is particularly effective when the analytical form of the objective function is unknown, non-differentiable, and computationally expensive to evaluate \cite{GARRIDOMERCHAN202020}. Therefore, we employ a Bayesian optimizer to solve the optimization problem defined in Equation~\ref{eq:optimal_problem}.

To evaluate the generalization of the optimized parameters, we first divide the OODEval-Human dataset into score intervals ([1,10], [11,20], …, [91,100]) according to human ratings. We then perform stratified sampling \cite{10.1007/978-3-030-75765-6_27} to construct training and test sets with an 8:2 split, following the practice in \cite{Gholamy2018Why7O,Bichri2024}, in order to reduce potential bias caused by uneven score distributions. The optimization initializes with uniform weights and terminates when no new optimum is found over 50 consecutive iterations, an empirically derived threshold that ensures convergence while minimizing computational overhead. Table~\ref{tab:correlation_on_train_test} presents the Pearson correlation coefficients ($\rho$) for the optimal parameters on both training and test sets ($\rho > 0.5$ indicates strong correlation, $\rho > 0.3$ moderate, and $\rho > 0.1$ weak~\cite{cohen1992statistical}), along with statistical significance ($p < 0.05$ denotes significance~\cite{tenny2017statistical}).

As shown in Table~\ref{tab:correlation_on_train_test}, all metrics demonstrate at least moderate correlations ($\rho > 0.3$) with human scores on the training set, while exhibiting robust generalization to the test set. In particular, the clue-class and clue metrics achieve strong correlations ($\rho > 0.5$). The p-values for all correlations are substantially below 0.05, indicating that the observed associations are statistically significant and reliable. The resulting optimal weight configuration is: $w_e = 0.810$, $w_r = 0.190$, $w_n = 0.787$, $w_a = 0.104$, $w_m = 0.109$, $w_{at} = 0.594$, $w_{an} = 0.406$, $w_{mn} = 0.730$, $w_{mt} = 0.153$, $w_{mp} = 0.117$, $w_{pt} = 0.050$, $w_{pn} = 0.950$, $w_{rt} = 0.156$, $w_{rm} = 0.220$, $w_{rn} = 0.624$. This configuration will serve as the baseline for subsequent empirical evaluations.

\subsection{Metrics Performance Evaluation}\label{sec:metrics_evaluation}

In the performance evaluation of our proposed metrics, we selected BLEU \cite{papineni2002bleu}, CodeBERTScore \cite{zhou-etal-2023-codebertscore}, and NINHS \cite{6933505} as baseline methods for comparison with CLUE. These baselines represent text matching, embedding-based, and domain-specific evaluation approaches, respectively. To assess their efficacy, we computed correlation coefficients between each automatic metric and human ratings derived from the OODEval-Human dataset. A higher correlation coefficient signifies stronger alignment between the automated metric and human preferences, thereby indicating superior metric quality \cite{10.1016/j.jss.2023.111741, zhou-etal-2023-codebertscore, xu2024human}. Table~\ref{tab:correlation_coefficient} reports the Pearson and Spearman correlation coefficients for the baselines, CLUE, and its sub-metrics in relation to human ratings, including their statistical significance levels.

The experimental results reveal two key insights: (1) CLUE substantially surpasses BLEU, CodeBERTScore, and NINHS in both Pearson and Spearman correlations with human ratings, with achieving high statistical significance. In particular, CLUE attains a Pearson correlation coefficient exceeding 0.5, which denotes a strong correlation with human judgments and bolsters interpretability \cite{cohen1992statistical}. (2) Relative to its constituent sub-metrics (i.e., clue-class, clue-attribute, clue-method, and clue-relation), the overarching clue metric exhibits superior correlations. This finding underscores that a holistic evaluation encompassing class names, attributes, methods, and relationships aligns more closely with human assessments and offers greater interpretability than evaluations focused on isolated elements.

\subsection{Ablation Study on the Underlying Semantic Model}\label{sec:ablation_study_on_semantic}

Since CLUE incorporates a modular semantic model for computing string similarities at the leaf nodes, it is essential to investigate the impact of varying underlying semantic models on its overall performance. To this end, we conducted an ablation study evaluating CLUE across multiple semantic models. Table~\ref{tab:ablation_study} presents the Pearson correlation coefficients achieved using four representative models, namely MiniLM \cite{NEURIPS2020_3f5ee243}, BGE \cite{bge_m3}, BERT \cite{devlin-etal-2019-bert}, and CodeBERT \cite{feng-etal-2020-codebert} for the CLUE metrics. The results indicate that the choice of underlying semantic model exerts a relatively minor influence on clue-attribute and clue-relation, but a more pronounced effect on clue-method and, particularly, clue-class. Specifically, the range of variation (i.e., the difference between the maximum and minimum correlations) for clue-attribute and clue-relation does not exceed 0.05; for clue-method, it is approximately 0.07; and for clue-class, it reaches about 0.186. Overall, the CLUE metrics grounded in CodeBERT deliver the highest performance across the board. Although the clue-relation and clue-attribute sub-metrics under CodeBERT are marginally inferior to those using BGE, the discrepancies are negligible, with gaps not surpassing 0.025. Consequently, owing to its superior holistic efficacy, we select CodeBERT as the underlying semantic model for all subsequent empirical investigations.

\section{Empirical Study}\label{sec:empirical_study}

Using OODEval and OODEval-Human, we evaluate existing LLMs and human on object-oriented design tasks to answer the following research questions. 

\begin{description}[leftmargin=0em, labelindent=0em]
  \item[• \textbf{RQ1 (Overall Correctness):}] How effective do LLMs perform on OOD tasks?
  \item[• \textbf{RQ2 (Comparison with Human):}] Do LLMs outperform undergraduates on OOD tasks?
  \item[• \textbf{RQ3 (Model Dimension Analysis):}] How do the different model dimensions affect the performance of LLMs on OOD tasks?
  \item[• \textbf{RQ4 (Task Feature Analysis):}] Which task features impact the LLMs' performance on OOD tasks?
  \item[• \textbf{RQ5 (Bad Case Analysis):}] What are the prevalent failure modes of LLMs on OOD tasks?
\end{description}

\begin{table}[h]
\vspace{-3mm}
\setlength{\tabcolsep}{9pt}
\caption{Studied LLMs.}
\vspace{-2mm}
\label{tab:model-info}
\begin{threeparttable}
\begin{tabular}{@{\extracolsep{\fill}} llccccccccc @{\extracolsep{\fill}}}
\toprule
& \textbf{Model} & \textbf{CL} & \textbf{Local?} & \textbf{OS?} & \textbf{CL?} & \textbf{IFT?} & \textbf{RLM?} & \textbf{PT} & \textbf{TDCD} & \\
\midrule
& llama3.1-8b-it & 128K & \CheckmarkBold & \CheckmarkBold & \XSolidBrush & \CheckmarkBold & \XSolidBrush & 2024-07 & 2023-12 & \\
& llama3.1-70b-it & 128K & \CheckmarkBold & \CheckmarkBold & \XSolidBrush & \CheckmarkBold & \XSolidBrush & 2024-07 & 2023-12 &\\
& llama2-7b & 4K & \CheckmarkBold & \CheckmarkBold & \XSolidBrush & \XSolidBrush & \XSolidBrush & 2023-07 & 2022-09 & \\
& llama2-13b & 4K & \CheckmarkBold & \CheckmarkBold & \XSolidBrush & \XSolidBrush & \XSolidBrush & 2023-07 & 2022-09 &\\
& codellama-7b-it & 16K & \CheckmarkBold & \CheckmarkBold & \CheckmarkBold & \CheckmarkBold & \XSolidBrush & 2023-08 & 2022-09 &\\
& codellama-13b-it & 16K & \CheckmarkBold & \CheckmarkBold & \CheckmarkBold & \CheckmarkBold & \XSolidBrush & 2023-08 & 2022-09 &\\
& codellama-34b-it & 16K & \CheckmarkBold & \CheckmarkBold & \CheckmarkBold & \CheckmarkBold & \XSolidBrush & 2023-08 & 2022-09 &\\
& codellama-7b & 16K & \CheckmarkBold & \CheckmarkBold & \CheckmarkBold & \XSolidBrush & \XSolidBrush & 2023-08 & 2022-09 &\\
& codellama-13b & 16K & \CheckmarkBold & \CheckmarkBold & \CheckmarkBold & \XSolidBrush & \XSolidBrush & 2023-08 & 2022-09 &\\
& codellama-34b & 16K & \CheckmarkBold & \CheckmarkBold & \CheckmarkBold & \XSolidBrush & \XSolidBrush & 2023-08 & 2022-09 &\\
& starcoder2-3b & 16K & \CheckmarkBold & \CheckmarkBold & \CheckmarkBold & \XSolidBrush & \XSolidBrush & 2024-03 & 2023-09 &\\
& starcoder2-7b & 16K & \CheckmarkBold & \CheckmarkBold & \CheckmarkBold & \XSolidBrush & \XSolidBrush & 2024-03 & 2023-09 &\\
& starcoder2-15b & 16K & \CheckmarkBold & \CheckmarkBold & \CheckmarkBold & \XSolidBrush & \XSolidBrush & 2024-03 & 2023-09 &\\
& gemma3-4b-it & 128K & \CheckmarkBold & \CheckmarkBold & \XSolidBrush & \CheckmarkBold & \XSolidBrush & 2025-03 & 2024-08 &\\
& gemma3-12b-it & 128K & \CheckmarkBold & \CheckmarkBold & \XSolidBrush & \CheckmarkBold & \XSolidBrush & 2025-03 & 2024-08 &\\
& gemma3-27b-it & 128K & \CheckmarkBold & \CheckmarkBold & \XSolidBrush & \CheckmarkBold & \XSolidBrush & 2025-03 & 2024-08 &\\
& gemma3-4b-pt & 128K & \CheckmarkBold & \CheckmarkBold & \XSolidBrush & \XSolidBrush & \XSolidBrush & 2025-03 & 2024-08 &\\
& gemma3-12b-pt & 128K & \CheckmarkBold & \CheckmarkBold & \XSolidBrush & \XSolidBrush & \XSolidBrush & 2025-03 & 2024-08 &\\
& gemma3-27b-pt & 128K & \CheckmarkBold & \CheckmarkBold & \XSolidBrush & \XSolidBrush & \XSolidBrush & 2025-03 & 2024-08 &\\
& qwen3-8B & 32K & \CheckmarkBold & \CheckmarkBold & \XSolidBrush & \CheckmarkBold & \CheckmarkBold & 2025-04 & 2024-11 &\\
& qwen3-14b & 32K & \CheckmarkBold & \CheckmarkBold & \XSolidBrush & \CheckmarkBold & \CheckmarkBold & 2025-04 & 2024-11 &\\
& qwen3-32b & 32K & \CheckmarkBold & \CheckmarkBold & \XSolidBrush & \CheckmarkBold & \CheckmarkBold & 2025-04 & 2024-11 &\\
& qwen3-coder-30b & 256K & \CheckmarkBold & \CheckmarkBold & \CheckmarkBold & \CheckmarkBold & \XSolidBrush & 2025-07 & 2024-11 &\\
& deepseek-coder-33b & 16K & \CheckmarkBold & \CheckmarkBold & \CheckmarkBold & \CheckmarkBold & \XSolidBrush & 2023-11 & 2023-03 &\\ 
& deepseek-r1 & 32K & \XSolidBrush & \CheckmarkBold & \XSolidBrush & \XSolidBrush & \CheckmarkBold & 2025-01 & 2024-07 &\\
& deepseek-v3 & 128K & \XSolidBrush & \CheckmarkBold & \XSolidBrush & \XSolidBrush & \XSolidBrush & 2024-07 & 2024-07 &\\
& gpt-3.5-turbo & 4K & \XSolidBrush & \XSolidBrush & \XSolidBrush & - & \XSolidBrush & 2024-03 & 2022-01 &\\
& gpt-4o-mini & 128K & \XSolidBrush & \XSolidBrush & \XSolidBrush & - & \CheckmarkBold & 2024-07 & 2023-10 &\\
& gpt-4o & 128K & \XSolidBrush & \XSolidBrush & \XSolidBrush & - & \CheckmarkBold & 2024-05 & 2023-10 &\\
\bottomrule
\end{tabular}
\begin{tablenotes}[flushleft]
  \item \textbf{CL}=Context Length; \textbf{OS}=Open Source; \textbf{CLM}=Code LLM; \textbf{IFT}=Instruct Fine-Tuned; \textbf{RLM}=Reasnoing LLM; \textbf{PT}=Publish Time; \textbf{TDCD}=Training Data Cutoff Date;
\end{tablenotes}
\vspace{-2mm}
\end{threeparttable}
\end{table}

\subsection{Studied LLMs}\label{sec:studied_llms}

To comprehensively assess the performance of  LLMs in OOD tasks, we evaluated 29 state-of-the-art LLMs across 8 distinct dimensions, building on recent empirical studies in software engineering tasks~\cite{11029885,10707668,wang2024oop}. As presented in Table~\ref{tab:model-info}, our study incorporates a diverse set of LLMs, characterized by diversity across the following dimensions: (1) \textbf{Model Parameter Scale}: encompassing local models with parameter scale ranging from 3 billion (starcoder2-3b~\cite{lozhkov2024starcoder2stackv2})  to 70 billion (Llama-3.1-70b-it~\cite{grattafiori2024llama3herdmodels}), alongside larger-scale online models. (2) \textbf{Context Length}: including models with context windows spanning from 4K to 128K tokens. (3) \textbf{Model Deployment}: covering both locally deployed and API-based online models(e.g., Deepseek series \cite{deepseekai2025deepseekr1incentivizingreasoningcapability,deepseekai2025deepseekv3technicalreport}, GPT series \cite{achiam2023gpt}). (4) \textbf{Model Accessibility}: comprising both open-source and closed-source models. (5) \textbf{Model Type}: encompassing code-specific models tailored for programming tasks(e.g., CodeLlama \cite{rozière2024codellamaopenfoundation}, Deepseek-Coder\cite{guo2024deepseekcoderlargelanguagemodel}) and general-purpose models designed for broader applications (e.g., Gemma3 series \cite{gemmateam2025gemma3technicalreport}). (6) \textbf{Model Training}: including instruction-fine-tuned models optimized for specific tasks and base models without task-specific fine-tuning. (7) \textbf{Reasoning Capability}: incorporating models that employ an explicit "thinking" process (e.g., the Qwen3 series~\cite{yang2025qwen3technicalreport}) and models lacking such features. (8) \textbf{Release Date}: covering models released between 2023 and 2025 to reflect recent advancements.

\subsection{Prompt Design}\label{sec:prompt_design}
In this section, we describe the prompt design used to enable LLMs to perform each OOD task in the OODEval benchmark. To ensure fair comparison across models, we follow prior benchmarking practices~\cite{10.1145/3597503.3639219,11029911} and adopt a unified prompt template applicable to both instruction-tuned and base LLMs. The template consists of a system role prompt that specifies the model’s assumed role and overall task goal, followed by task-specific instructions detailing the requirement description and expected output. Models are instructed to generate PlantUML code enclosed by \textbf{@startuml} and \textbf{@enduml} markers to facilitate automatic extraction, with explicit inclusion of class names, attributes, methods, and inter-class relationships.

\begin{promptbox}[title=Prompt for Object-Oriented Design]
\#\# System Message \\
you are a professional UML class diagram design expert, which can generate the corresponding PlantUML class diagram design based on system requirements. \\
\#\# Instruction \\
System requirement is as follows :\\
$\{user\_requirement\}$ \\
please generate plantuml code based on system requirement. You should follow the instructions below: 
\begin{enumerate}
    \item Generate standard PlantUML class diagram code directly, start with \textit{@startuml} and end with \textit{@enduml} tags. Do not generate any analysis, explanations, or irrelevant content.
    \item Class names, attributes, and method names should use meaningful english names from system requirement.
    \item Reasonably use inheritance, implementation, dependence, association, aggregation, and composition relationships to design the class diagram.
\end{enumerate} 
\#\# Response
\end{promptbox}

\subsection{Grammatical Correctness Evaluation Metric}

In addition to using the proposed CLUE metrics to evaluate the semantic correctness of the generated designs, we also adopt the widely used Pass@k metric~\cite{chen2021evaluatinglargelanguagemodels} to assess grammatical correctness, which quantifies the proportion of successful outcomes across \(k\) generated samples per task. In the context of code generation, this metric assesses the percentage of test cases that pass based on \(k\) generated code samples. Similarly, we adapt it here to measure the syntax pass rate for \(k\) generated design examples, using the PlantUML parser as an oracle to verify correctness.
\begin{equation}
     \text{Pass@k} = \mathop{\mathbb{E}}_{\text{Designs}} \left[ 1 - \binom{n - c}{k}/\binom{n}{k} \right]
     \label{eq:pass@k}
\end{equation}
In Equation \ref{eq:pass@k}, \(n\) denotes the total number of design samples generated, \(c\) represents the number of correct samples that successfully pass validation by the PlantUML parser, and \(k\) corresponds to the number of samples considered in the Pass@k calculation.

\subsection{Implementation Details}

For large-scale online LLMs, such as GPT-4o and GPT-3.5-turbo, we accessed the models via their respective API interfaces. For open-source local LLMs, we downloaded and executed their released versions directly from the official repositories, following the provided documentation. To ensure consistency, the maximum context window length was uniformly set to 4096 tokens across all models, corresponding to the smallest supported window among the evaluated LLMs. All experiments were conducted on a server equipped with 8 NVIDIA L40S GPUs, each with 46 GB of memory. For each task, we generated five samples of design solutions, yielding a total of 7,250 outputs (29 LLMs × 50 tasks × 5 samples). Following prior studies~\cite{chen2021evaluatinglargelanguagemodels,10.1145/3597503.3639219,10.1145/3691620.3695470}, we adopted a temperature of 0.2, default values for top-$p$ and top-$k$, and a maximum output token length of 2048 to produce complete responses without truncation.

\section{RESULTS}

\subsection{RQ1: Overall Correctness}\label{sec:rq1}

\begin{table}[t]
\vspace{-2mm}
\caption{Performance of Large Language Models on OODEval: \textit{pass@1} and \textit{CLUE} Metrics.}
\vspace{-2mm}
\label{tab:model-metrics}
\resizebox{\textwidth}{!}{
{\small
\begin{tabular}{@{}llccccccc@{}}
\toprule
 & \textbf{Model} & \textbf{pass@1} & \textbf{clue} & \textbf{clue-class} & \textbf{clue-attribute} & \textbf{clue-method} & \textbf{clue-relation} & \\ \midrule
 & llama3.1-8b-it  & .844 $\pm$ .056  & .735 $\pm$ .046  & .761 $\pm$ .050  & .730 $\pm$ .051  & .537 $\pm$ .083  & .627 $\pm$ .036  & \\
 & llama3.1-70b-it  & .984 $\pm$ .008  & .850 $\pm$ .017  & .885 $\pm$ .020  & .864$^*$ $\pm$ .025  & .686 $\pm$ .069  & .702 $\pm$ .016  & \\
 & llama2-7b  & .048$^\dag$ $\pm$ .010  & .022$^\dag$ $\pm$ .003  & .025$^\dag$ $\pm$ .004  & .026$^\dag$ $\pm$ .004  & .018$^\dag$ $\pm$ .003  & .008$^\dag$ $\pm$ .001  & \\
 & llama2-13b  & .452 $\pm$ .076  & .282 $\pm$ .041  & .340 $\pm$ .058  & .344 $\pm$ .058  & .197 $\pm$ .061  & .037 $\pm$ .007  & \\
 & codellama-7b-it  & .848 $\pm$ .055  & .603 $\pm$ .046  & .687 $\pm$ .059  & .638 $\pm$ .053  & .475 $\pm$ .092  & .244 $\pm$ .031  & \\
 & codellama-13b-it  & .984 $\pm$ .008  & .745 $\pm$ .026  & .799 $\pm$ .028  & .782 $\pm$ .026  & .500 $\pm$ .107  & .515 $\pm$ .054  & \\
 & codellama-34b-it  & .972 $\pm$ .014  & .712 $\pm$ .023  & .765 $\pm$ .030  & .722 $\pm$ .033  & .495 $\pm$ .100  & .485 $\pm$ .047  & \\
 & codellama-7b  & .864 $\pm$ .039  & .604 $\pm$ .036  & .649 $\pm$ .043  & .587 $\pm$ .040  & .421 $\pm$ .078  & .415 $\pm$ .033  & \\
 & codellama-13b  & .932 $\pm$ .030  & .685 $\pm$ .032  & .722 $\pm$ .036  & .670 $\pm$ .048  & .469 $\pm$ .092  & .525 $\pm$ .042  & \\
 & codellama-34b  & .940 $\pm$ .020  & .666 $\pm$ .036  & .721 $\pm$ .040  & .658 $\pm$ .046  & .481 $\pm$ .106  & .430 $\pm$ .038  & \\
 & starcoder2-3b  & .616 $\pm$ .094  & .430 $\pm$ .060  & .456 $\pm$ .066  & .453 $\pm$ .063  & .252 $\pm$ .051  & .319 $\pm$ .042  & \\
 & starcoder2-7b  & .900 $\pm$ .040  & .680 $\pm$ .046  & .709 $\pm$ .049  & .668 $\pm$ .048  & .553 $\pm$ .068  & .559 $\pm$ .053  & \\
 & starcoder2-15b  & 1.00$^*$ $\pm$ .000  & .815 $\pm$ .024  & .845 $\pm$ .025  & .808 $\pm$ .023  & .776$^*$ $\pm$ .038  & .687 $\pm$ .034  & \\
 & gemma3-4b-it  & 1.00$^*$ $\pm$ .000  & .814 $\pm$ .018  & .815 $\pm$ .022  & .803 $\pm$ .021  & .586 $\pm$ .084  & .810 $\pm$ .015  & \\
 & gemma3-12b-it  & .996 $\pm$ .001  & .840 $\pm$ .016  & .844 $\pm$ .019  & .797 $\pm$ .022  & .607 $\pm$ .100  & .825$^*$ $\pm$ .018  & \\
 & gemma3-27b-it  & 1.00$^*$ $\pm$ .000  & .843 $\pm$ .020  & .847 $\pm$ .025  & .818 $\pm$ .024  & .697 $\pm$ .071  & .823 $\pm$ .017  & \\
 & gemma3-4b-pt  & .392 $\pm$ .082  & .261 $\pm$ .049  & .266 $\pm$ .051  & .238 $\pm$ .044  & .188 $\pm$ .038  & .241 $\pm$ .049  & \\
 & gemma3-12b-pt  & .856 $\pm$ .035  & .698 $\pm$ .041  & .710 $\pm$ .043  & .682 $\pm$ .040  & .526 $\pm$ .086  & .645 $\pm$ .044  & \\
 & gemma3-27b-pt  & .980 $\pm$ .005  & .787 $\pm$ .025  & .813 $\pm$ .026  & .802 $\pm$ .025  & .590 $\pm$ .080  & .678 $\pm$ .034  & \\
 & qwen3-8b  & .720 $\pm$ .050  & .575 $\pm$ .041  & .584 $\pm$ .044  & .556 $\pm$ .046  & .395 $\pm$ .064  & .537 $\pm$ .037  & \\
 & qwen3-14b  & .872 $\pm$ .036  & .716 $\pm$ .041  & .737 $\pm$ .043  & .704 $\pm$ .042  & .545 $\pm$ .073  & .628 $\pm$ .040  & \\
 & qwen3-32b  & .936 $\pm$ .020  & .787 $\pm$ .031  & .801 $\pm$ .033  & .763 $\pm$ .030  & .591 $\pm$ .085  & .728 $\pm$ .029  & \\
 & qwen3-coder-30b  & .980 $\pm$ .007  & .864$^*$ $\pm$ .019  & .886$^*$ $\pm$ .022  & .849 $\pm$ .023  & .774 $\pm$ .051  & .770 $\pm$ .019  & \\
 & deepseek-coder-33b  & .988 $\pm$ .004  & .789 $\pm$ .027  & .786 $\pm$ .030  & .768 $\pm$ .030  & .624 $\pm$ .062  & .801 $\pm$ .024  & \\
 & deepseek-r1  & 1.00$^*$ $\pm$ .000  & .863 $\pm$ .014  & .881 $\pm$ .015  & .813 $\pm$ .020  & .674 $\pm$ .063  & .787 $\pm$ .018  & \\
 & deepseek-v3  & .972 $\pm$ .011  & .854 $\pm$ .020  & .867 $\pm$ .021  & .803 $\pm$ .032  & .708 $\pm$ .058  & .799 $\pm$ .024  & \\
 & gpt-3.5-turbo  & .996 $\pm$ .001  & .810 $\pm$ .023  & .836 $\pm$ .025  & .772 $\pm$ .024  & .633 $\pm$ .071  & .702 $\pm$ .030  & \\
 & gpt-4o-mini  & 1.00$^*$ $\pm$ .000  & .807 $\pm$ .024  & .837 $\pm$ .026  & .776 $\pm$ .025  & .642 $\pm$ .066  & .677 $\pm$ .031  & \\
 & gpt-4o  & 1.00$^*$ $\pm$ .000  & .856 $\pm$ .020  & .875 $\pm$ .021  & .819 $\pm$ .024  & .686 $\pm$ .063  & .775 $\pm$ .027  & \\ \bottomrule
\end{tabular}
}}
\vspace{-2mm}
\begin{flushleft}
\small
\textbf{*} denotes the maximum value; \textbf{\dag} denotes the minimum value.
\end{flushleft}
\vspace{-2mm}
\end{table}

\begin{figure*}[t]
    \centering
    \includegraphics[width=0.7\linewidth]{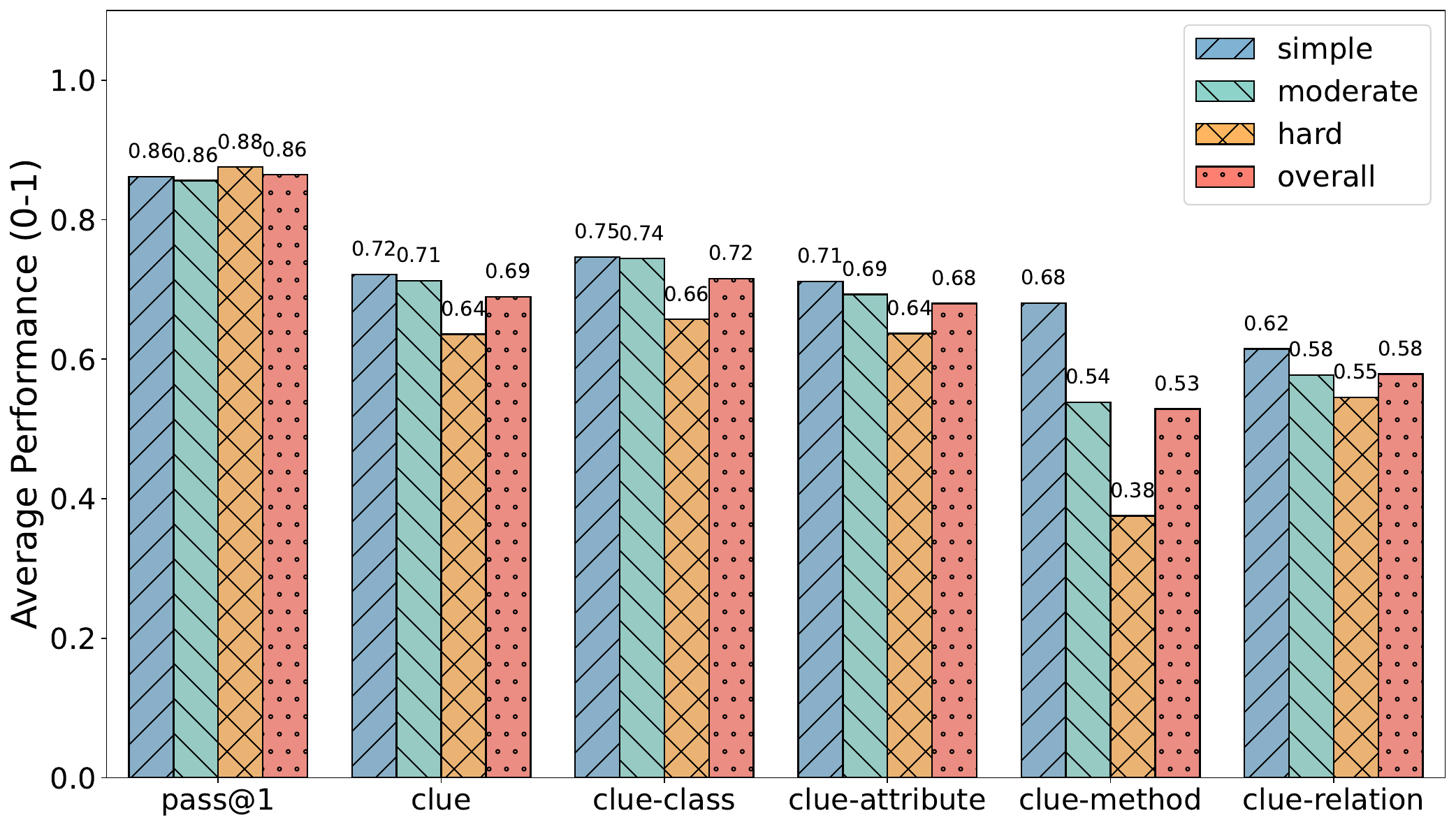}
    \caption{Performance Comparison of Pass@1 and CLUE Metrics.}
    \label{fig:compare_among_metrics}
\end{figure*}

\begin{figure*}[ht]
    \centering
    \includegraphics[width=1.0\linewidth]{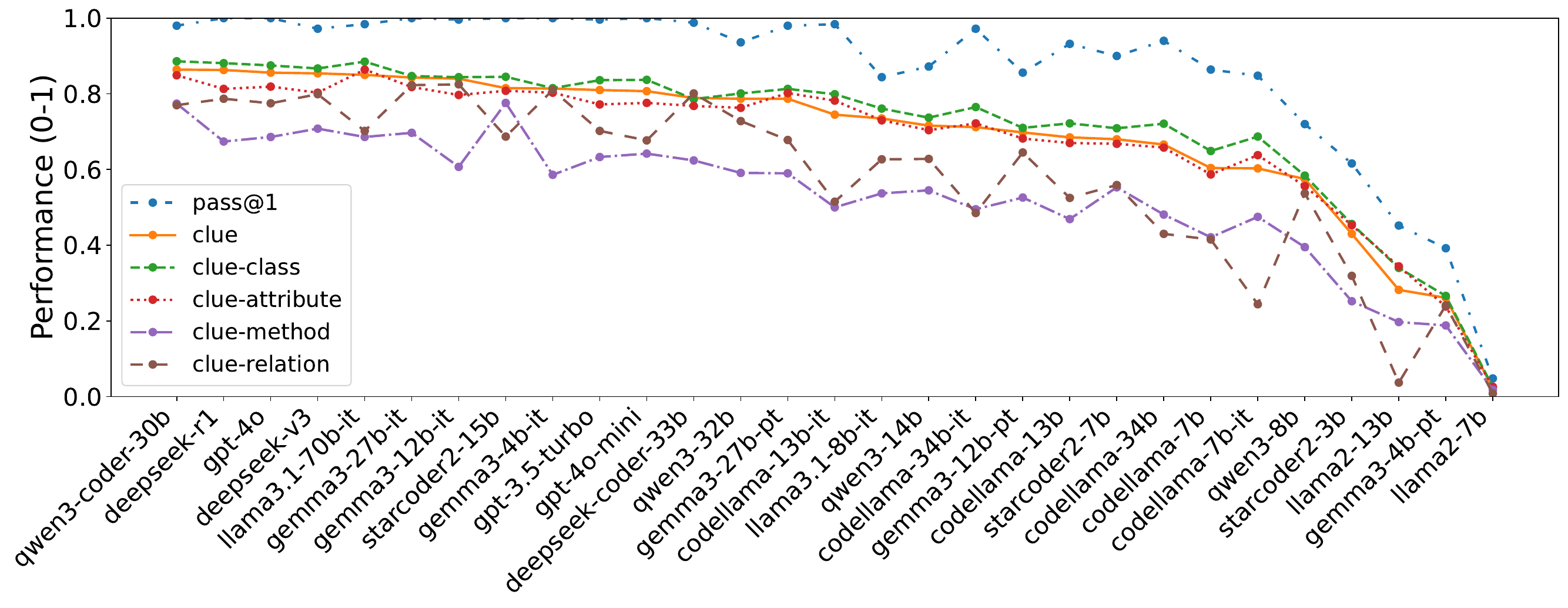}
    \caption{ Performance Ranking of LLMs Across Metrics (sorted by \textit{clue}). }
    \label{fig:performance_ranking}
\end{figure*}

In this section, we first present the evaluation results of all LLMs on grammatical correctness (measured by \textit{pass@1}) and semantic correctness (assessed via \textit{CLUE} metrics) using the OODEval dataset, as summarized in Table~\ref{tab:model-metrics}. To further examine the variations across metrics and difficulty levels, we report the LLMs' average performance of all metrics on the simple, moderate, and hard subsets, as well as on the entire dataset, as depicted in Figure~\ref{fig:compare_among_metrics}. Additionally, we illustrate the performance ranking of the LLMs based on their overall clue scores to provide an intuitive overview of the models' relative differences, as shown in Figure~\ref{fig:performance_ranking}. Based on Table~\ref{tab:model-metrics}, Figure~\ref{fig:compare_among_metrics}, and Figure~\ref{fig:performance_ranking}, we derive the following key observations.

In Figure~\ref{fig:compare_among_metrics}, all evaluated LLMs achieve an average \textit{pass@1} score exceeding 80\% across simple, moderate, hard, and overall sets. And in table~\ref{tab:model-metrics}, six models attaining a perfect \textit{pass@1} score of 100\% on the overall set. This demonstrates high syntactic correctness in generating PlantUML code. Nevertheless, \textit{CLUE} scores are consistently over 10\% lower than \textit{pass@1} scores across all difficulty levels, indicating relatively weak semantic correctness in the generated code.

Analysis of the \textit{CLUE} metrics in Figure~\ref{fig:compare_among_metrics} shows that, across datasets of varying difficulty levels as well as the overall dataset, \textit{clue-class} and \textit{clue-attribute} consistently outperform \textit{clue-method} and \textit{clue-relation}. Moreover, the performance gap becomes more pronounced as task difficulty increases. This indicates that LLMs are relatively effective at generating accurate class names and attributes, but exhibit clear deficiencies in synthesizing class methods and inter-class relationships. As defined in Section~\ref{subsubsec:clue_metrics}, the overall \textit{clue} score is composed of \textit{clue-class} and \textit{clue-relation}, where \textit{clue-class} is positively correlated with both \textit{clue-attribute} and \textit{clue-method}. Consequently, the weaker performance in class method and relationship generation substantially constrains the overall model performance.

Further examination of Figure~\ref{fig:compare_among_metrics} reveals that as task difficulty increases across the simple, moderate, and hard OODEval subsets, \textit{CLUE} metrics exhibit varying degrees of performance decline, with \textit{clue-method} showing the most pronounced decrease. In contrast, \textit{pass@1} scores remain relatively stable across difficulty levels. These obeservations indicate that syntactic correctness (\textit{pass@1}) is largely insensitive to task complexity, whereas semantic correctness (\textit{CLUE} metrics) is relatively sensitive, with LLMs generally exhibiting poorer semantic correctness on more complex tasks.

\begin{summary-rq}
\textbf{Finding 1:} 
LLMs exhibit strong syntactic correctness but show significant weaknesses in semantic correctness. Specifically, their performance in generating methods and relationships is notably poorer, particularly on the more challenging subsets of OODEval. Furthermore, task difficulty has a clear impact on semantic correctness, while syntactic correctness remains stable across different difficulty levels.
\end{summary-rq}

According to Table~\ref{tab:model-metrics} and Figure~\ref{fig:performance_ranking}, there is significant variability in performance on the OODEval benchmark. The lowest-performing model, LLaMA2-7B, achieves a score near 0, while the top performer, Qwen3-Coder30B, attains approximately 0.86. Over half of the evaluated models score above 0.7. Among the top 10 models, the performance gaps are narrow, within 6 percentage points. Notably, alongside high-performing online models (DeepSeek-V3, DeepSeek-R1, GPT-4o), local models Qwen3-Coder-30B and Llama3.1-70B-IT rank among the top 5, demonstrating competitive performance. Unexpectedly, GPT-4o-Mini ranks outside the Top 10, while local open-source models from the Gemma3-IT series and Starcoder2-15B outperform the closed-source online model GPT-4o-Mini, with all entering the Top 10 (Gemma3-27B-IT, Gemma3-12B-IT, Gemma3-4B-IT, and Starcoder2-15B are ranked 6th, 7th, 8th, and 9th, respectively). These findings indicate that, in OOD tasks, local open-source models can rival online closed-source models.

Furthermore, we observe that different models exhibit distinct strengths and weaknesses across classes, attributes, methods, and relationships. Specifically, LLaMA3.1-70B-IT achieves the highest performance on \textit{clue-attribute}, with a score of 0.864; Starcoder2-15B performs best on \textit{clue-method}, reaching 0.776; and Gemma3-12B-IT excels in \textit{clue-relation}, with a top score of 0.825. However, Gemma3-12B-IT performs relatively poorly on \textit{clue-method} but still ranks highly overall due to its strong performance in \textit{clue-relation}. In contrast, although LLaMA3.1-70B-IT leads in \textit{clue-attribute}, it ranks lower overall than Qwen3-Coder-30B because of its weaker performance in \textit{clue-method} and \textit{clue-relation}. Qwen3-Coder-30B achieves the best overall \textit{clue} score by exhibiting more balanced performance across attributes, methods, and relationships, even though it does not achieve the highest score in any individual sub-metric. These findings highlight that LLMs possess complementary strengths in generating class attributes, methods, and inter-class relationships, and suggest that balanced capabilities across these dimensions are critical for achieving strong overall performance on OOD tasks.

\begin{summary-rq}
\textbf{Finding 2:} 
The Qwen3-Coder-30B demonstrates exceptional and consistent performance, surpassing other models in the benchmark. The local LLaMA3.1-70B-IT model performs comparably to online models such as DeepSeek-V3, DeepSeek-R1, and GPT-4o, ranking in the second tier. Interestingly, the smaller-scale Gemma3-4B-IT outperforms GPT-4o-Mini, underscoring the potential for optimizing models under resource constraints. Additionally, models demonstrate imbalanced performance across class attributes, methods, and relationships, while top-tier models tend to achieve a more balanced performance in these dimensions.
\end{summary-rq}

\subsection{RQ2: Comparison With Human}

\begin{figure*}[h]
    \centering
    \includegraphics[width=0.7\linewidth]{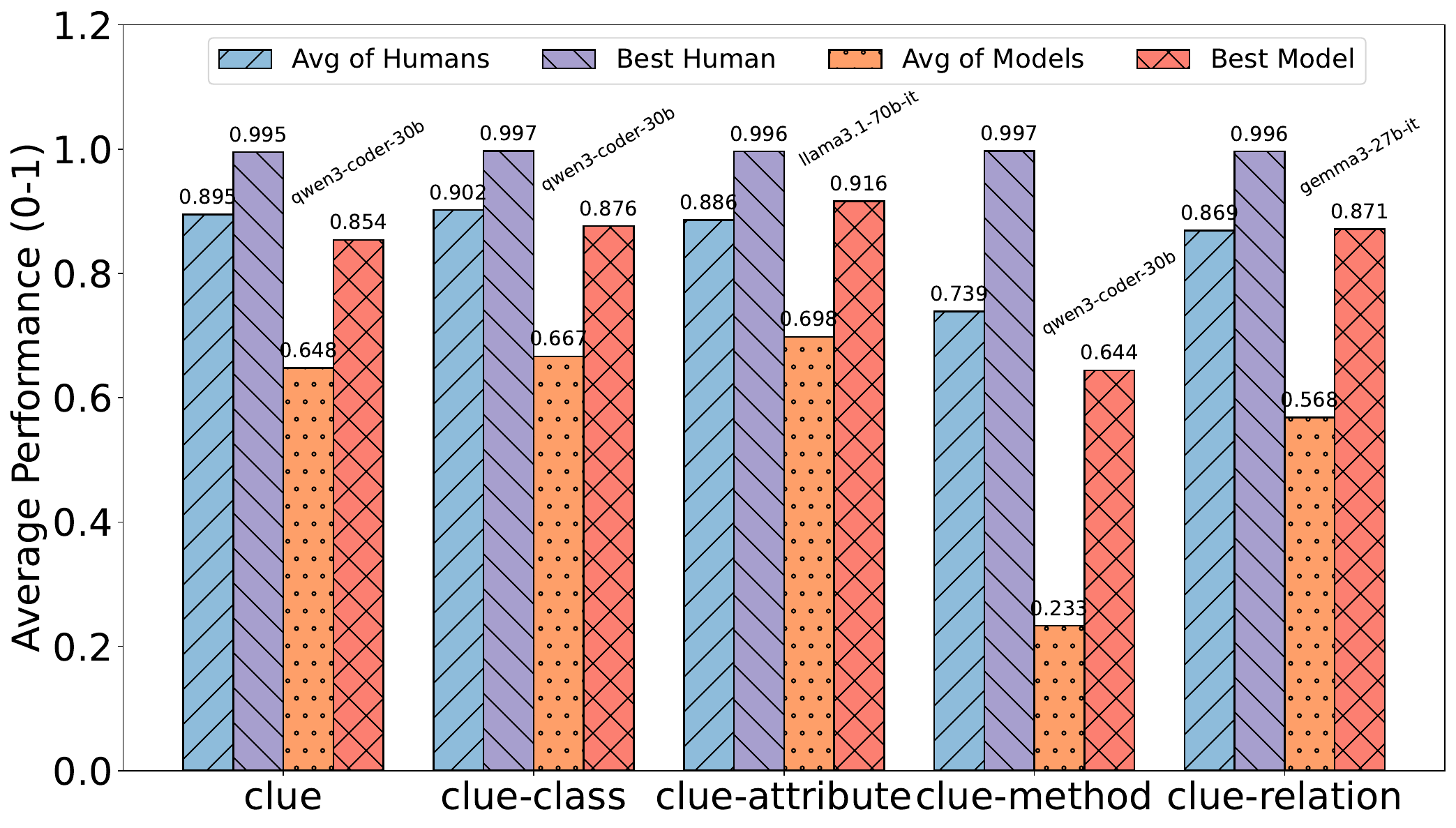}
    \caption{Performance Comparison of LLMs and Human.}
    \label{fig:human_models_bar_chart}
\end{figure*}

The OODEval-Human dataset contains design solutions completed by undergraduate students. To enable a direct and equitable comparison between LLMs and human participants in object-oriented design scenarios, we re-evaluate each solution in OODEval-Human using the CLUE metrics, which ensure measurement consistency with that used for LLMs. As illustrated in Figure~\ref{fig:human_models_bar_chart}, which presents the average and peak performance of LLMs and humans across the \textit{CLUE} metrics, we derive the following key observations.

First, comparing the average performance of humans to that of LLMs reveals that LLMs significantly underperform the human average across all \textit{CLUE} metrics. The most pronounced disparities occur in class method generation (\textit{clue-method}, exceeding a 50\% gap) and class relationship identification (\textit{clue-relation}, exceeding a 30\% gap). Collectively, these results indicate that the average LLMs remain substantially inferior to the average human proficiency in OOD tasks. Second, when benchmarking the human average against the best LLM performance, notable findings emerge. The top model for attribute generation (Llama3.1-70B-Instruct) marginally surpasses the human average, while the leading model for relationship generation (Gemma3-27B-Instruct) approximates it closely. In contrast, the strongest performer across the overall \textit{clue}, \textit{clue-class}, and \textit{clue-method} metrics (Qwen3-Coder-30B) falls slightly below the human average. These deviations are minimal, not exceeding 5 percentage points. Thus, the premier LLMs achieve near parity with average human performance. Finally, contrasting the peak human performance with the best LLMs highlights a clear disparity. The highest-performing human participants accomplish OOD tasks with near-perfect accuracy, attaining scores above 99\% on all \textit{CLUE} metrics. Evidently, even the most advanced LLMs currently exhibit a considerable gap relative to the best human benchmark.

\begin{summary-rq}
\textbf{Finding 3:} 
In object-oriented design (OOD) tasks, the average performance of LLMs falls substantially below that of average human participants. However, state-of-the-art LLMs have approached or, in certain metrics, slightly surpassed the human average. Nevertheless, a considerable gap remains between the best-performing LLMs and peak human performance.
\end{summary-rq}

\subsection{RQ3: Model Dimension Analysis}\label{sec:dimension_analysis}

\begin{figure*}[t]
    \centering
    \begin{subfigure}[t]{0.66\textwidth}
        \centering
        \includegraphics[width=\linewidth]{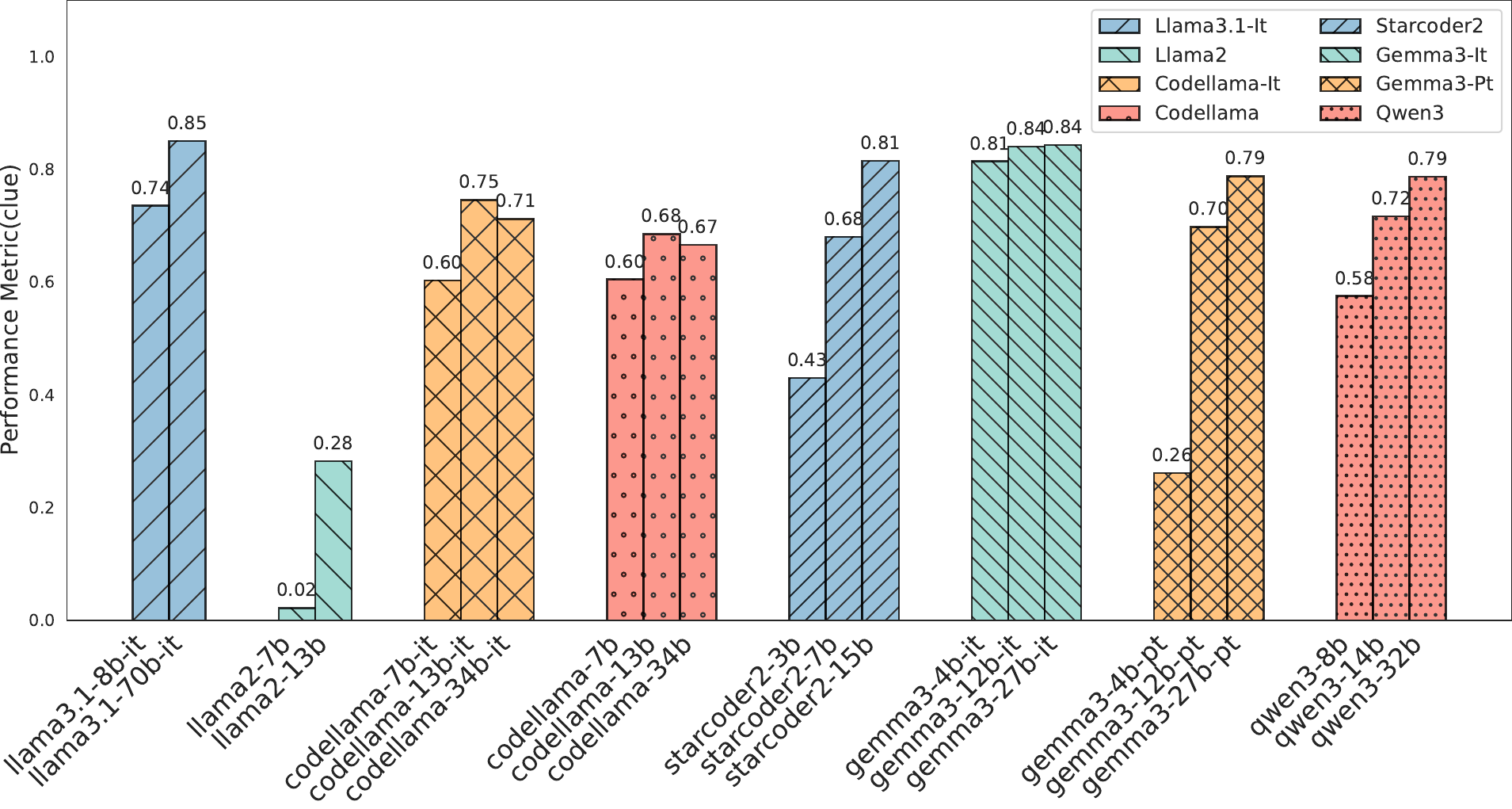}
        \caption{Model Parameter Scale.}
        \label{fig:model_dimension_analysis_1}
    \end{subfigure}%
    \hfill
    \begin{subfigure}[t]{0.32\textwidth}
        \centering
        \includegraphics[width=\linewidth]{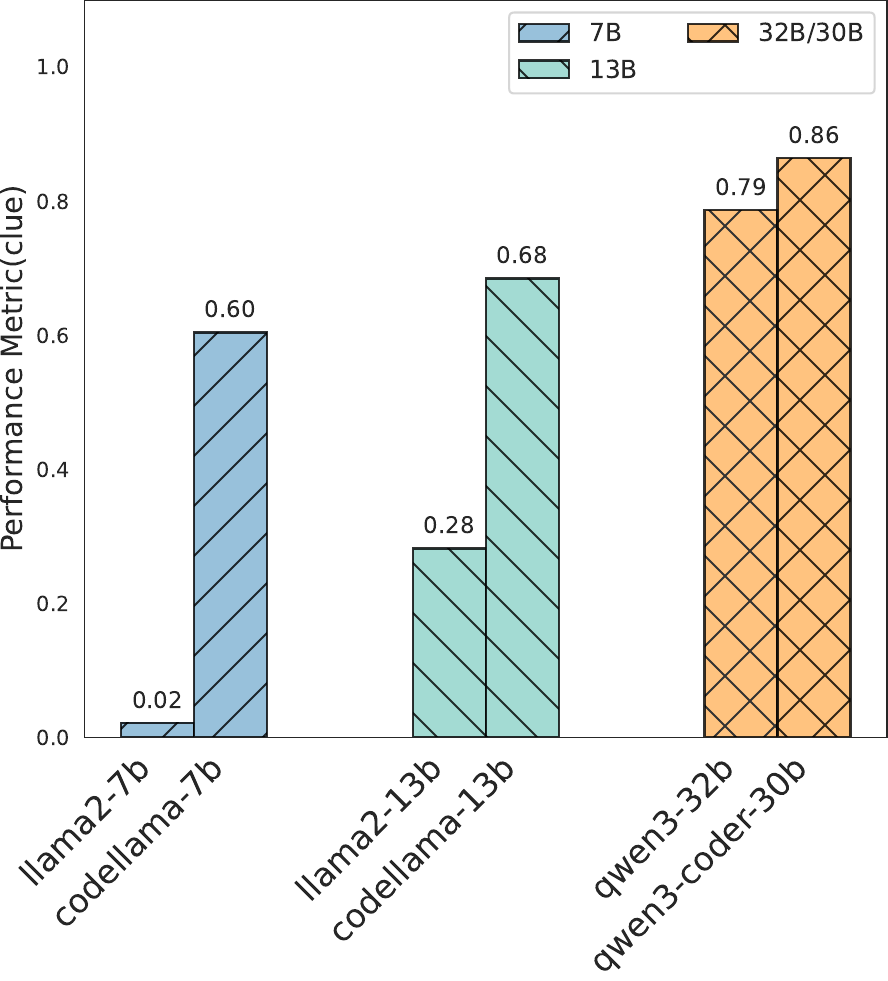}
        \caption{Code LLMs vs General LLMs.}
        \label{fig:model_dimension_analysis_2}
    \end{subfigure}

    \vspace{1em}  

    \begin{subfigure}[t]{0.64\textwidth}
        \centering
        \includegraphics[width=\linewidth]{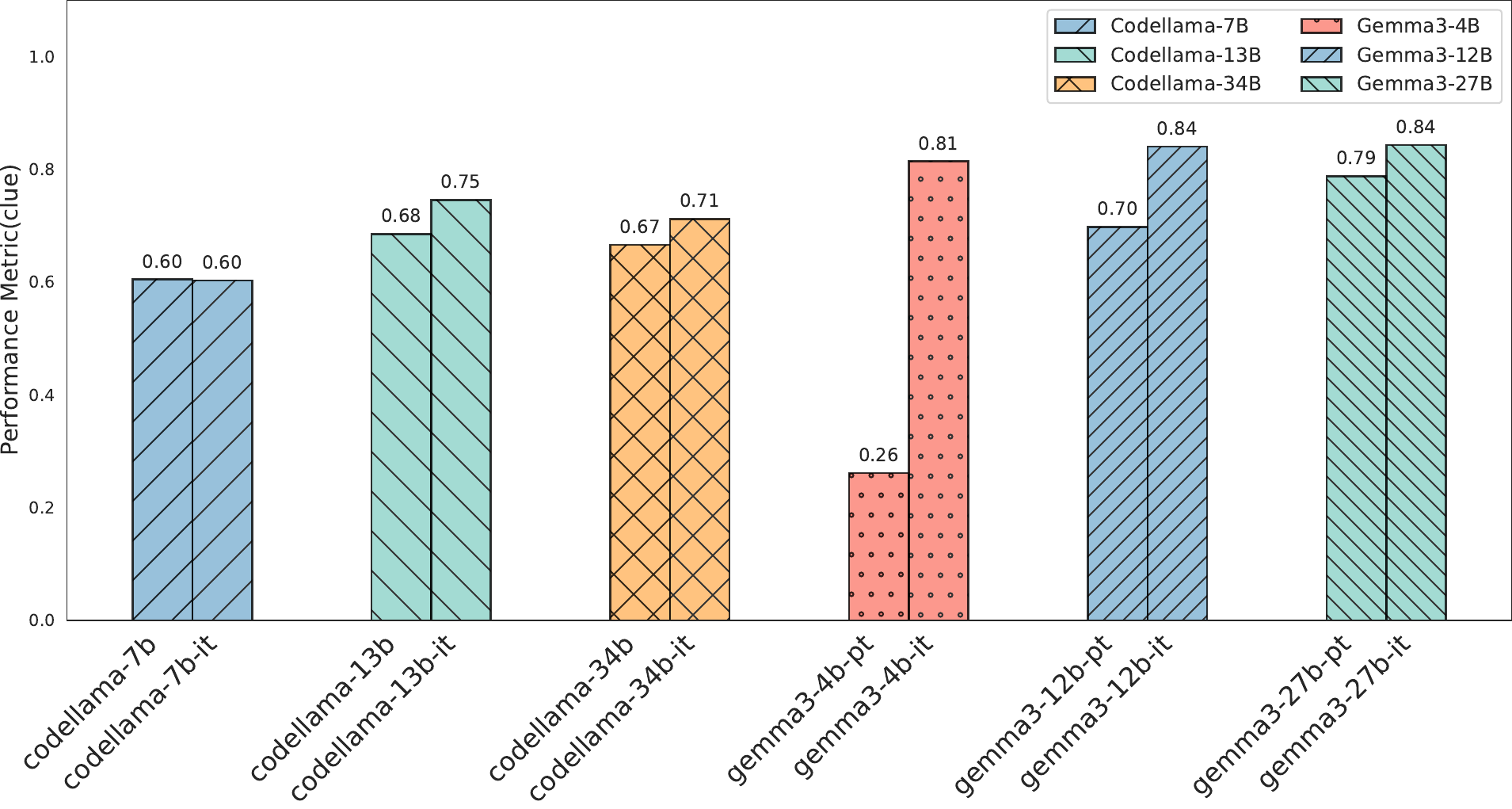}
        \caption{Base LLMs vs Instruct Fine-tuned LLMs.}
        \label{fig:model_dimension_analysis_3}
    \end{subfigure}%
    \hfill
    \begin{subfigure}[t]{0.34\textwidth}
        \centering
        \includegraphics[width=\linewidth]{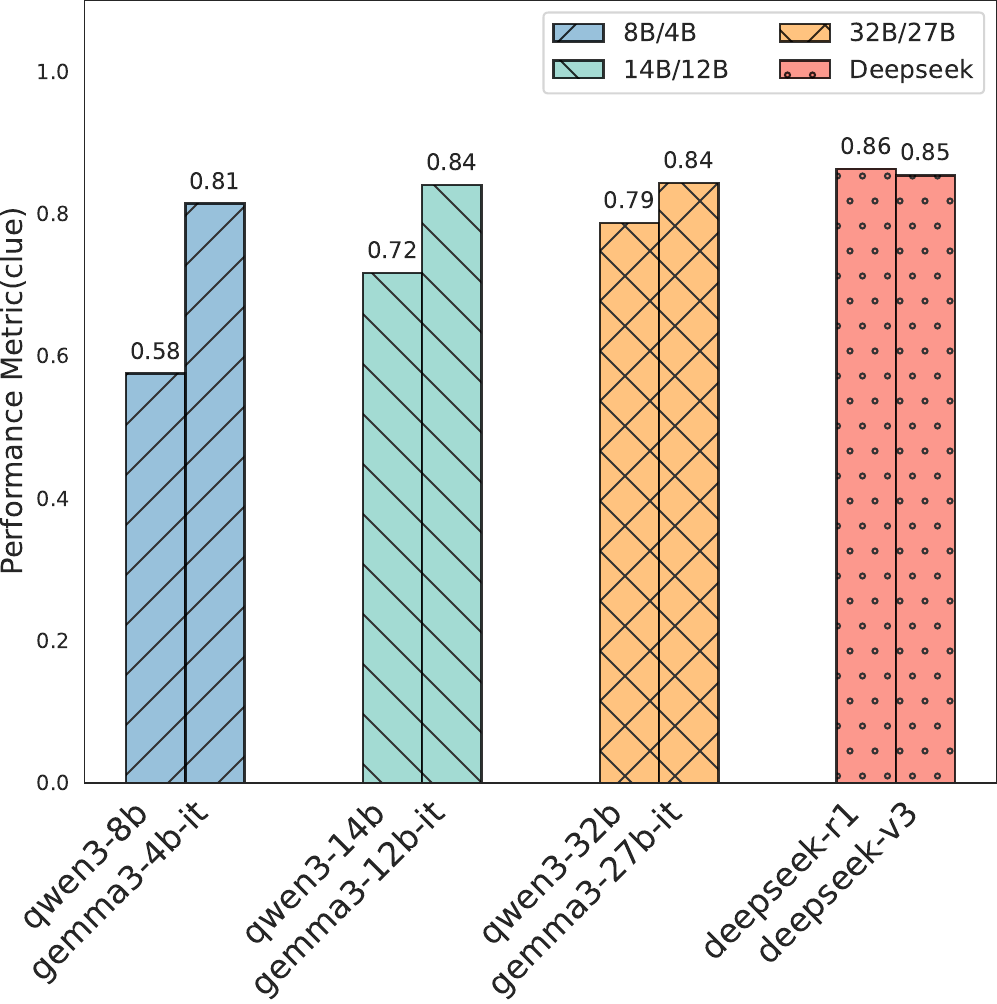}
        \caption{RLMs vs non-RLMs.}
        \label{fig:model_dimension_analysis_4}
    \end{subfigure}

    \caption{Model Performance Analysis Across Four Dimensions.}
    \label{fig:model_dimension_analysis}
\end{figure*}

Based on the model classifications in Table~\ref{tab:model-info}, we analyzed four comparison groups using the OODEval benchmark: (1) model parameter scale, (2) base LLMs versus instruction-tuned LLMs, (3) code-specific LLMs versus general-purpose LLMs, and (4) reasoning-enhanced LLMs versus non-reasoning LLMs (see Figure~\ref{fig:model_dimension_analysis}).

To assess the impact of model parameter scale, we selected eight groups of models sharing the same architecture but varying in parameter size, as shown in Figure~\ref{fig:model_dimension_analysis_1}. The results show that performance generally improves with increasing parameter count, except for minor declines in CodeLlama-34B-Instruct relative to CodeLlama-13B-Instruct and CodeLlama-34B relative to CodeLlama-13B. This pattern likely arises from larger models' superior comprehension of task requirements, enabling more accurate extraction of class-related elements.

To compare code-specific LLMs with general-purpose LLMs, we selected three model pairs, as illustrated in Figure~\ref{fig:model_dimension_analysis_2}. CodeLlama derives from Llama2 via code dataset training, forming one control group, while Qwen3-32B and Qwen3-Coder-30B form another due to their same model series and similar scales. Results indicate that code-specific models significantly outperform general-purpose ones, suggesting that code-focused secondary training enhances OOD task performance. This may occur because PlantUML code resembles programming languages like Python in syntax, allowing code-specific models to generate superior PlantUML outputs.

To evaluate instruction tuning's impact, we selected six pairs of base models and their instruction-tuned counterparts from the same series and parameter scale, as shown in Figure~\ref{fig:model_dimension_analysis_3}. Instruction-tuned versions consistently outperform base models across all pairs, with the Gemma3-4B tuned variant showing over 50\% improvement. These results affirm instruction tuning's predictable and robust efficacy on OOD tasks.

Finally, we compared reasoning LLMs(RLMs) with non-reasoning LLMs(non-RLMs) using four model pairs, as depicted in Figure~\ref{fig:model_dimension_analysis_4}. Qwen3 (released April 2025) is reasoning-enhanced, while Gemma3 (released March 2025) is not. We paired Qwen3 and Gemma3 models of comparable parameter scales for comparison. Additionally, DeepSeek-R1 and DeepSeek-V3 were included as another control pair. The results reveal that Gemma3 outperforms Qwen3, and DeepSeek-V3 trails DeepSeek-R1 by just 1 percentage point. Thus, reasoning models show no clear advantage and may underperform compared to their non-reasoning counterparts. Since we did not ensure identical underlying model architectures and model parameter scale across pairs, this finding should be regarded as preliminary observations.

\begin{summary-rq}
\textbf{Finding 4:} 
Larger model parameter scales generally enhance performance on OOD tasks; code-specific LLMs significantly outperform general-purpose LLMs; and instruction-tuned models consistently surpass their base counterparts, achieving substantial improvements. In contrast, reasoning LLMs exhibit no clear advantage over non-reasoning LLMs. Overall, parameter scale, code specialization, and instruction tuning emerge as key factors influencing OOD task performance.
\end{summary-rq}

\subsection{RQ4: Task Feature Analysis}\label{sec:feature_analysis}

\begin{figure*}[ht]
    \centering
    \begin{subfigure}[t]{0.57\textwidth}
        \centering
        \includegraphics[width=\linewidth]{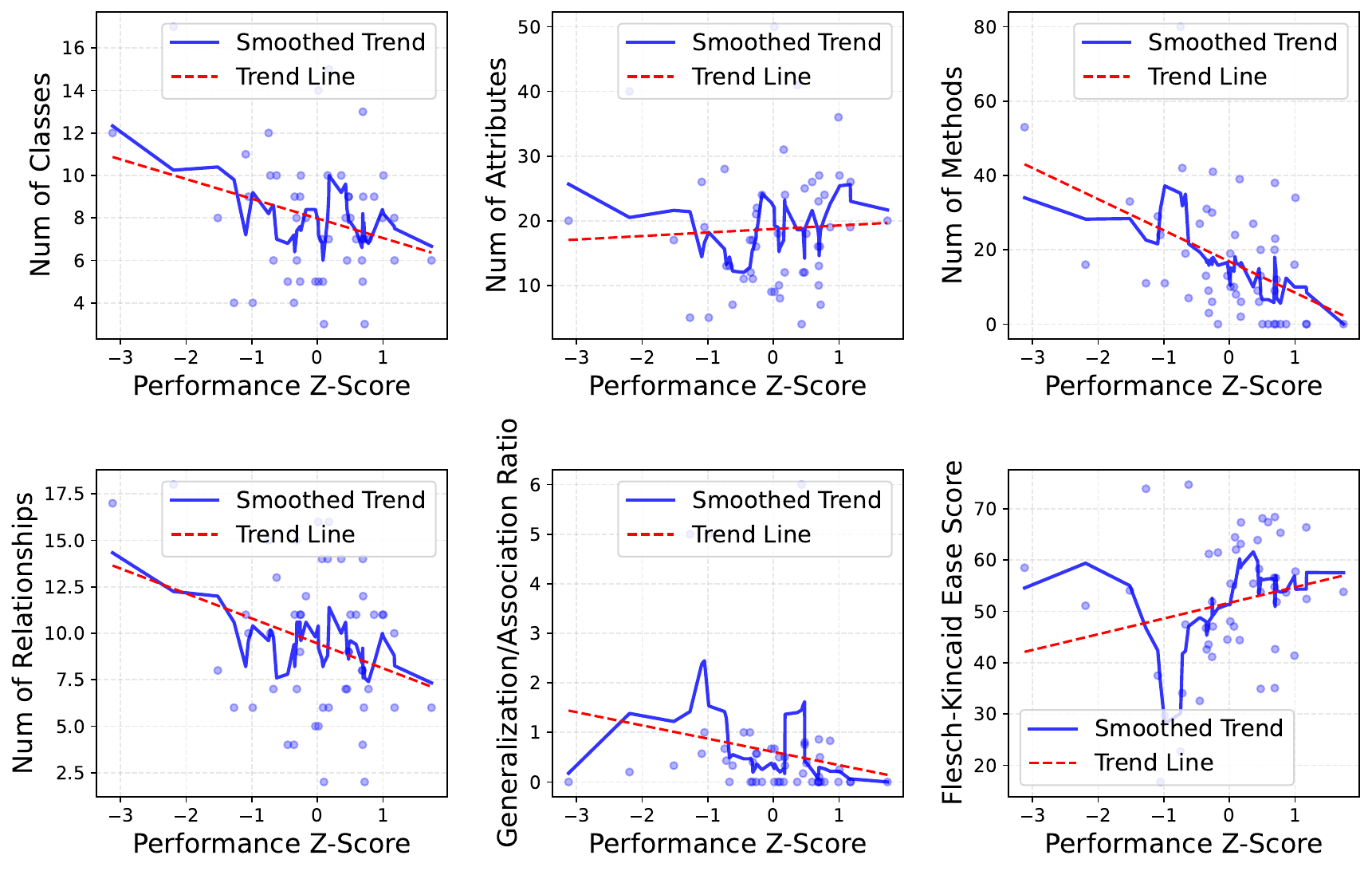}
        \caption{Feature Trends vs. Performance Z-Score.}
        \label{fig:feature_trends}
    \end{subfigure}
    \hfill
    \begin{subfigure}[t]{0.42\textwidth}
        \centering
        \includegraphics[width=\textwidth]{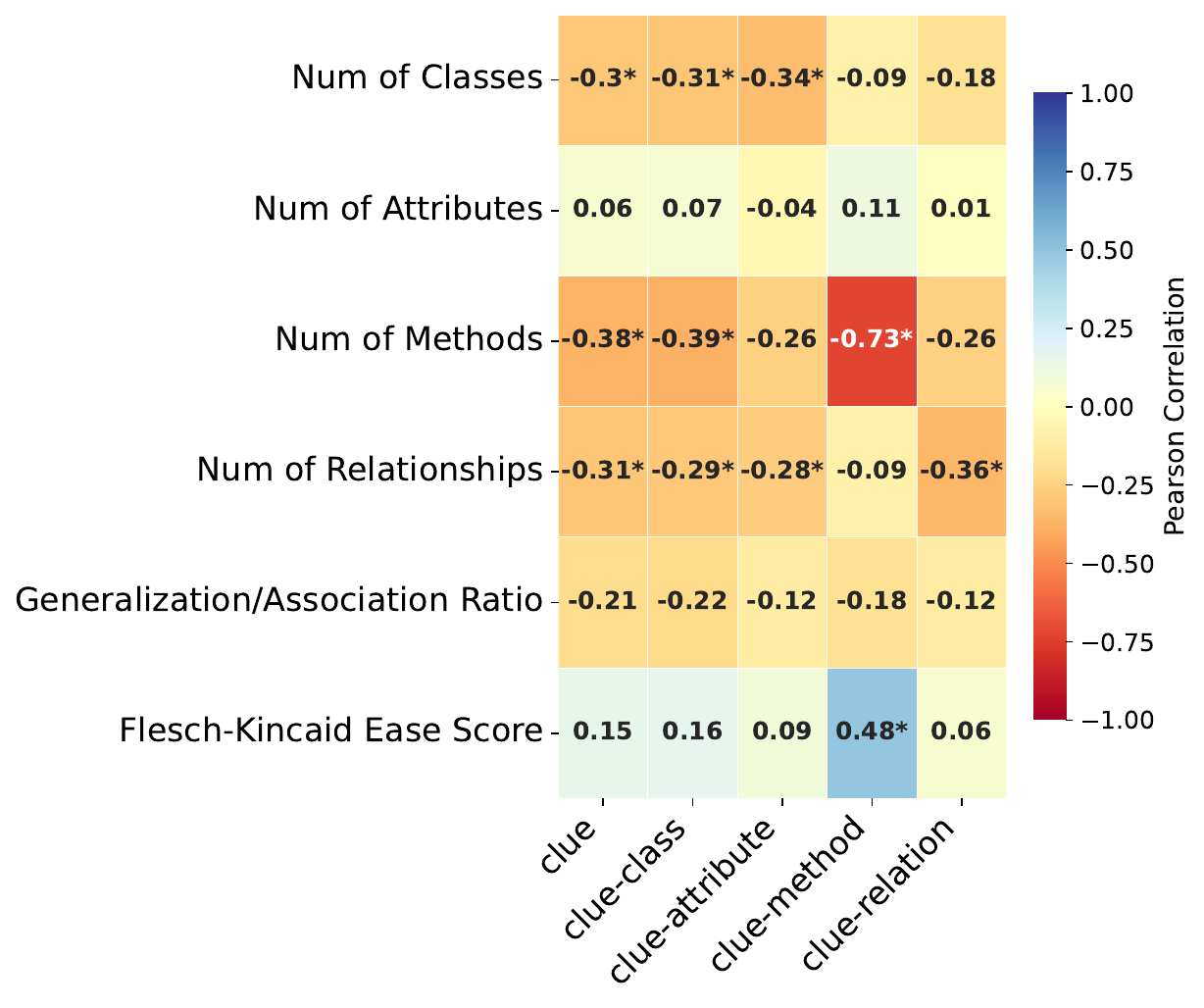}
        \caption{Feature-Metric Correlation Matrix.}
        \label{fig:feature_correlation}
    \end{subfigure}
    \vspace{-4mm}
    \caption{Analysis of Task Features that Affect the Performance of LLMs.} 
    \label{fig:feature_analysis}
\end{figure*}

Figure~\ref{fig:feature_trends} depicts the performance trends across six key task features: the number of classes, the number of attributes, the number of methods, the number of relationships, the ratio of generalization relationships to association relationships, and the Flesch-Kincaid Reading Ease Score. To capture these variations, we first apply Z-score normalization to the CLUE metrics across 29 LLMs and compute their mean values. Figure~\ref{fig:feature_correlation} further presents the Pearson correlation coefficients and statistical significance levels for these six features with respect to each CLUE metric. Notably, correlation coefficients marked with an asterisk indicate a significance level of $p > 0.05$. Based on these figures, we derive the following observations.

From Figure~\ref{fig:feature_trends}, we observe that LLM performance consistently declines as the number of classes, methods, relationships, and the generalization-to-association ratio increase. This suggests that more complex class structures, extensive method sets, and denser relationship, particularly those involving generalization relationships (e.g., subclass inheritance or interface realization), pose substantial challenges for LLMs. Conversely, higher Flesch-Kincaid Reading Ease Scores correlate with improved performance, implying that reduced text readability impairs model capabilities. Additionally, performance remains largely stable with an increasing number of attributes, aligning with our findings in Section~\ref{sec:rq1} that attribute generation is not a primary bottleneck for these models.

To identify which task features most influence specific metrics, we further examine the correlations and significance levels between the six features and each CLUE metric. As illustrated in Figure~\ref{fig:feature_correlation}, the number of attributes exhibits no correlation with any metric, reinforcing our earlier observations. The generalization-to-association ratio shows weak correlations across all metrics, with no statistical significance. In contrast, the number of relationships demonstrates significant correlations with all CLUE metrics except CLUE-Method, underscoring that relationship generation, encompassing both generalization and association types, represents a clear bottleneck, though association relationships appear slightly less challenging than generalization ones. The Flesch-Kincaid Reading Ease Score exhibits a significant correlation with CLUE-Method but weak or negligible correlations with others, indicating that lower readability complicates the extraction and identification of operations from requirements. In summary, the number of classes, methods, and relationships each show significant correlations with at least three metrics, highlighting their pivotal role in determining LLM performance.

\begin{summary-rq}
\textbf{Finding 5:} 
LLM performance drops with increases in the number of classes, methods, relationships, and generalization-to-association ratios, as well as with lower readability scores, while the number of attributes shows no effect. The numbers of classes, methods, and relationships are the primary bottlenecks, exhibiting significant correlations with most CLUE metrics, whereas the generalization-to-association ratio and readability have more selective influences.
\end{summary-rq}

\subsection{RQ5: Bad Case Analysis}

We classified all failure cases into three categories: Instruction Failure, Syntax Error, and Semantic Bias. Instruction Failure occurs when the model deviates from the prompt instructions (Section~\ref{sec:prompt_design}) for generating PlantUML code delimited by @startuml and @enduml, preventing successful code extraction. Syntax Error refers to cases where the extracted code contains syntactic violations. Semantic Bias denotes instances where the extracted code is syntactically valid but deviates from the reference diagram in meaning or completeness. Our analysis reveals the following distribution across all generation results: Instruction Failure at 7.08\% (affecting 11 LLMs), Syntax Error at 6.69\% (affecting 22 LLMs), and Semantic Bias at 86.23\% (affecting all 29 LLMs).

\begin{figure*}[t]
    \vspace{-5mm}
    \centering
    \includegraphics[width=1.0\linewidth]{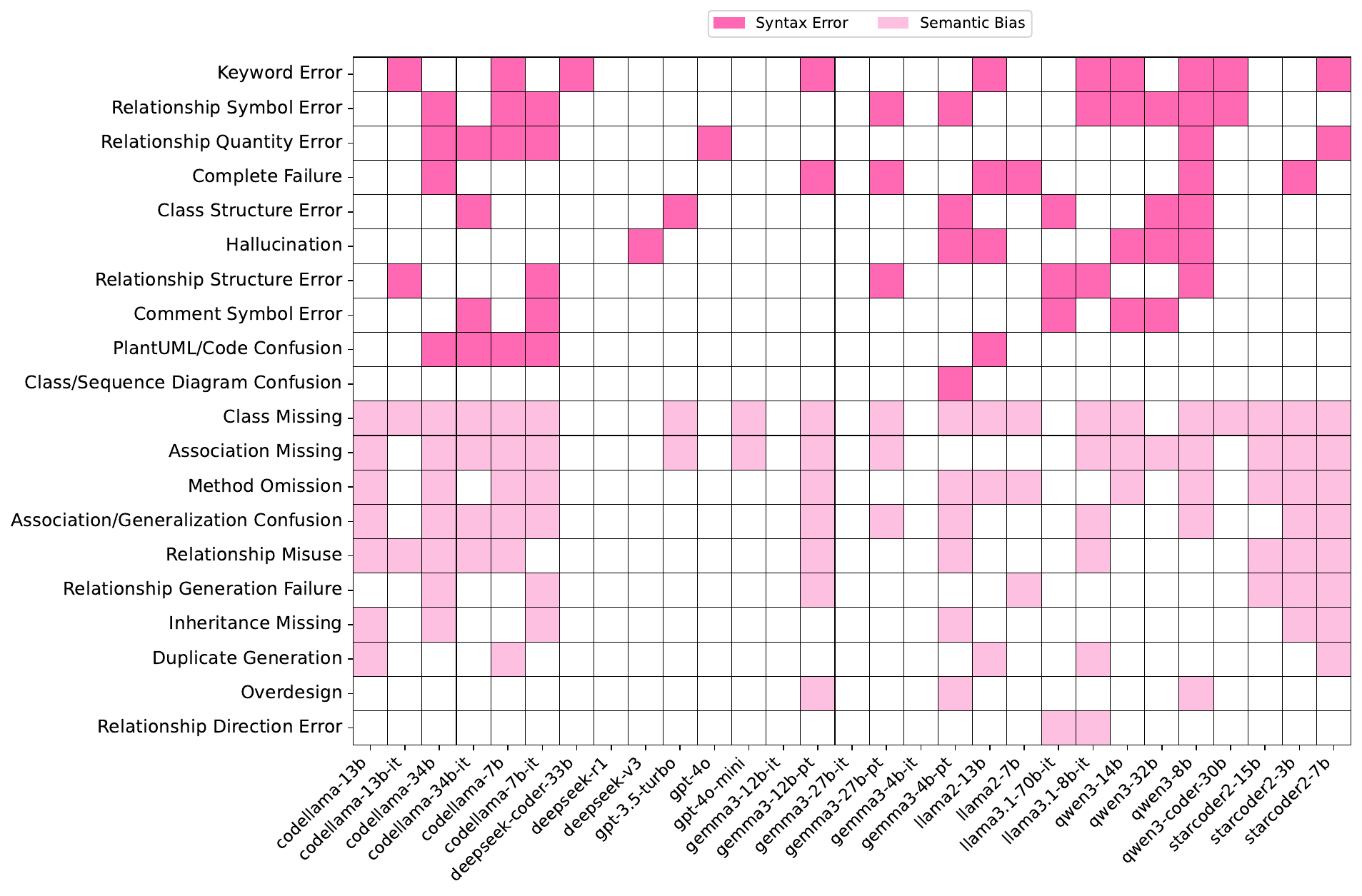}
    \vspace{-9mm}
    \caption{The Prevalent Failure Modes Distributed in Bad Cases Across All Models.}
    \label{fig:bad_case_empty_low}
    \vspace{-5mm}
\end{figure*}

Given the resource-intensive nature of examining all 7,250 generation results, we randomly sampled 10\% (725 results) for in-depth analysis. Focusing on Syntax Error and Semantic Bias cases, we reviewed their output logs and identified 10 common issue types per category. These encompass nine issues related to class relationships, four to class elements, five to structural aspects, and two to miscellaneous types. Instruction Failure cases were omitted due to the absence of extractable code. Figure~\ref{fig:bad_case_empty_low} depicts the co-occurrence of these 20 issue types across the 29 LLMs, grouped into seven series: CodeLlama, DeepSeek, GPT, Gemma, LLaMA, Qwen, and StarCoder. Error types are ordered by their prevalence across models.

To address potential ambiguities, we define key error types as follows: Complete Failure involves generating irrelevant content despite proper @startuml and @enduml delimiters. Hallucination entails adding extraneous elements beyond the requirements, often causing syntax errors. PlantUML/Code Confusion arises when the model outputs class implementation code instead of diagram syntax. Method Omission occurs when class methods are excluded from the diagram. Relationship Misuse reflects an over-reliance on a single relationship type (e.g., aggregation, composition, or inheritance). Duplicate Generation involves redundant replication of class elements, such as attributes or relationships. Overdesign results in unnecessarily decomposing attributes into separate classes, increasing complexity.

As shown in Figure~\ref{fig:bad_case_empty_low}, prevalent issues like Keyword Error, Class Missing, and Association Missing span six model series, underscoring their ubiquity. Moderately common issues, appearing in four to five series, include Relationship Symbol Error, Relationship Quantity Error, Complete Failure, Class Structure Error, Hallucination, Relationship Structure Error, Method Omission, Association/Generalization Confusion, Relationship Misuse, and Relationship Generation Failure. Model-specific errors, limited to one to three series, encompass PlantUML/Code Confusion (predominant in CodeLlama), Comment Symbol Error, Class/Sequence Diagram Confusion (unique to Gemma3-4B-PT), Inheritance Missing, Duplicate Generation, Overdesign (observed in three models), and Relationship Direction Error (unique to LLaMA 3.1).

Evaluating error diversity across the 29 LLMs, models such as CodeLlama-34B, CodeLlama-7B, CodeLlama-7B-IT, Gemma3-4B-PT, Qwen3-8B, and StarCoder2-7B exhibited 10–12 distinct error types, indicating high error diversity. Conversely, the DeepSeek series, GPT series, instruction-tuned (IT) variants of Gemma3, and Qwen3-Coder-30B displayed only 1–3 error types, signifying low diversity. The remaining models showed 4–9 error types, reflecting moderate diversity. Notably, DeepSeek-R1, Gemma3-12B-IT, Gemma3-4B-IT, and Gemma3-27B-IT exhibited no errors, demonstrating superior performance. Overall, the IT variants of Gemma3 not only achieved high overall correctness (Section~\ref{sec:rq1}) but also presented the lowest error diversity, positioning them as preferred models for future OOD tasks in this domain.

\begin{summary-rq}
\textbf{Finding 6:} 
Keyword Error, Class Missing, and Association Missing are the most common issues across different model families. Relationship Symbol Error, Hallucination, and Method Omission occur less frequently but are still notable. The DeepSeek series, GPT series, the IT version of Gemma3, and Qwen3-Coder-30b exhibit low diversity, encountering only a limited number of error types. In particular, the IT version of Gemma3 demonstrates superior overall performance, making it the ideal candidate for future local deployment in OOD research.
\end{summary-rq}

\section{Implication and Future Directions}

Based on the empirical findings of this study, this section explores the broader theoretical and practical implications of these results, with a particular focus on model improvement, model selection, benchmark expansion, and software engineering education, thereby providing significant guidelines for advancing future research and practical applications in the field.

\subsection{Implications for Model Improvement}

\textbf{Enhancing the model's understanding of operations and relationships.} The Finding 1 reveals significant limitations in the models' generation of class methods and class relationships. Furthermore, the Finding 5 indicates that model performance declines markedly as the number of methods and relationships increases. Therefore, targeted improvements in the model's comprehension of operational descriptions and relational semantics in system requirements will contribute to better overall performance in OOD tasks.

\textbf{Leveraging the complementary strengths of multiple models.} The Finding 2 demonstrates imbalances in the capabilities of different models across various design elements, with each exhibiting unique strengths; models with more balanced capabilities (e.g., Qwen3-Coder-30B) generally achieve superior overall performance. Without additional training, integrating the specialized strengths of multiple models in specific element generation (e.g., via routing mechanisms or ensemble strategies) holds promise for further enhancing OOD task performance.

\textbf{Fine-tuning large-scale code-specific LLMs.} The Finding 4 confirms that models with larger parameter scales, code-specific pre-training, and instruction fine-tuning significantly outperform baselines in OOD tasks. Accordingly, applying targeted fine-tuning to existing large-scale code LLMs (e.g., Qwen3-Coder-30B) offers considerable potential for performance improvement.

\subsection{Implications for Model Selection}

\textbf{Local deployment scenarios with sufficient resources.} For environments with ample computational resources, Qwen3-Coder-30B is recommended for local deployment. This model delivers the most optimal and balanced performance across all OOD task metrics(as shown in RQ1). It ranks first in overall correctness while exhibiting relatively low error diversity (as indicated in Finding 6), making it well-suited for high-performance local environments where both accuracy and consistency are paramount.

\textbf{Scenarios with limited training resources.}
In resource-constrained environments, Gemma3-4b-IT is the ideal base model for fine-tuning. Despite its smaller scale, this model demonstrates strong overall correctness (as shown in Finding 1), even outperforming GPT-4o-mini. Furthermore, Finding 6 reveals that it has the lowest diversity in error types, with minimal occurrences of syntactic errors or significant semantic deviations. This makes it an excellent choice for applications where resource efficiency is crucial without compromising on accuracy.

\textbf{Practical production deployment scenarios.} For production environments where cost-effectiveness and reliability are prioritized, the Deepseek-R1 API service is the preferred option. It ranks second in overall correctness, surpassing the GPT series, and maintains low error diversity. Additionally, it offers significant cost savings compared to GPT models, making it a strong candidate for applications that require a balance between performance and economic efficiency.

\subsection{Implications for Benchmark Expansion}

\textbf{Developing benchmarks with more difficult tasks.} The task features analysis indicates that the number of classes, methods, and relationships is a critical factor influencing model performance. Therefore, extending benchmark datasets to include more instances with large-scale classes, numerous methods, and complex relationships would better reflect real-world software design scenarios.

\textbf{Increasing the proportion of generalization relationships.} The task features analysis shows a performance decline as the proportion of generalization relationships increases, suggesting ongoing challenges in handling such relationships. Future benchmarks should intentionally increase the inclusion of tasks involving generalization relationships to more accurately identify model limitations.

\subsection{Implications for Software Engineering Education}

\textbf{Remain vigilant against student cheating and establish mechanisms to counter AI-assisted academic misconduct.} The human-model comparison in RQ2 demonstrates that state-of-the-art models achieve performance levels comparable to those of average undergraduate students, with only minor shortcomings in method implementation. This proficiency enables students to generate complete design solutions using these models, potentially enabling academic misconduct in assignments or exams while attaining moderate-to-high grades. To safeguard academic integrity, software engineering educators must adopt proactive countermeasures, such as emphasizing process-oriented assessments, requiring in-person demonstrations or oral defenses, and developing specialized AI plagiarism detection tools.

\textbf{Guide students in responsibly utilizing AI for software design exploration to enhance authentic learning outcomes.} Although AI tools introduce risks to academic integrity, they also offer valuable opportunities for educational enhancement. By redesigning curricula to prioritize critical thinking, innovation, and ethical AI integration, educators can encourage responsible model usage. This strategy not only mitigates misconduct risks but also empowers students to leverage AI as a supportive tool, thereby fostering deeper comprehension and genuine skill development in AI-assisted software engineering contexts.

\section{Threats to Validity}

\textbf{Threats to Internal Validity} concern the accuracy of our study's causal conclusions. One threat is errors in model implementations. To mitigate this, we used official model versions from HuggingFace. Another threat is data leakage between our OODEval benchmark and the training data of evaluated models. We addressed this by manually constructing OODEval and involving multiple participants to validate its integrity, reducing subjectivity and errors. Additionally, the choice of semantic model could skew evaluation metrics. To counter this, we conducted an ablation study on various semantic models and selected the one with the highest correlation to human evaluation scores. 

\textbf{Threats to External Validity} relate to the generalizability of our findings. The limited scale and single-reference design of OODEval may restrict its applicability to broader contexts. To improve generalizability, we plan to expand OODEval in future work by including a wider range of tasks and reference implementations.

\textbf{Threats to Construct Validity} address the suitability of our measurement methods. A potential bias in the OODEval-Human dataset could misrepresent metric performance. To ensure data authenticity and diversity, we collected diverse submissions from students across multiple academic years and task types. Another concern is model output truncation due to token length limits, which could affect PlantUML code extraction. We conducted a preliminary study to evaluate output lengths and set a maximum token limit of 4096, ensuring complete and accurate outputs.

\section{Conclusion}

This study represents the first attempt to evaluate the performance of 29 distinct LLMs alongside human undergraduates on OOD tasks. We first developed OODEval, a novel OOD benchmark dataset, and constructed OODEval-Human, the first huamn-rated benchmark including undergraduate OOD solutions with instructor ratings. Subsequently, we introduced the CLUE metrics to facilitate both global and fine-grained evaluations of design elements. Our empirical investigation encompassed five dimensions: overall performance, human-LLM comparisons, model dimension analysis, task feature analysis, and bad case analysis. Findings reveal pronounced deficiencies in LLMs' generation of class methods and relationships; Qwen3-Coder-30B exhibits superior performance and low error diversity, rendering it ideal for local deployment; LLMs' best performance nearly match the performance of average humans; Parameter scale, code specialization, and instruction tuning emerge as key factors influencing LLMs’ performanc; design complexity markedly impacts LLM efficacy; and prevalent LLM failure modes include keyword errors,  classes or associations missing, and method omssion. These insights offer valuable guidance for advancing LLM research, practical deployment, and educational strategies in software design.

\section*{Acknowledgments}

This work was supported by the Major Program of the National Natural Science Foundation of China (Grant Nos. 62192733, 62192730) and the National Natural Science Foundation of China (Grant No. 62372219).

\bibliographystyle{ACM-Reference-Format}
\bibliography{sample-base}

\end{document}